\documentclass{elsarticle}
\usepackage{amssymb,amsmath,graphicx}
\usepackage{lineno,hyperref}
\usepackage{tipa}
\usepackage{subcaption}
\modulolinenumbers[5]
\usepackage{stmaryrd}
\usepackage{algorithm}
\usepackage[noend]{algpseudocode}
\newcommand{\R}{{\mathbb{R}}}

\newcommand{\C}{\mathbb{C}}
\newcommand{\cplus}{\mathbb{C}^{+}}

\newcommand{\abs}[1]{\left | {#1} \right |}
\renewcommand{\v}[0]{\check}

\newcommand{\DisplayNote}[1]{\hspace{.5cm}({\bf{#1}})}

\journal{Measurement}

\begin{document}

\begin{frontmatter}

\title{An acoustic glottal source \\ for vocal tract physical models} %%
%% Group authors per affiliation:
\author[mymainadress]{Antti Hannukainen}
\author[mymainadress]{Juha Kuortti}
\author[mymainadress,mysecondaryaddress]{Jarmo Malinen\corref{mycorrespondingauthor}}
\author[mymainadress]{Antti Ojalammi}

\address[mymainaddress]{Dept. Mathematics and Systems Analysis, Aalto University, Finland}
\address[mysecondaryaddress]{Dept. Signal Processing and Acoustics, Aalto University, Finland}
\cortext[mycorrespondingauthor]{Corresponding author}
%\ead[url]{www.math.aalto.fi}
% \ead{support@elsevier.com}

\begin{abstract}
A sound source was proposed for acoustic measurements of physical
models of the human vocal tract. The physical models are produced by
Fast Prototyping, based on Magnetic Resonance Imaging during prolonged
vowel production. The sound source, accompanied by custom signal
processing algorithms, is used for two kinds of measurements:
(\textrm{i}) amplitude frequency response and resonant frequency
measurements of physical models, and (\textrm{ii}) signal
reconstructions at the source output according to a target waveform
with measurements at the mouth position of the physical model.  The
proposed source and the software are validated by measurements on a
physical model of the vocal tract corresponding to vowel
[\textipa{\textscripta}] of a male speaker.
\end{abstract}


\end{frontmatter}

%\linenumbers

\section{Introduction}

Acquiring comprehensive data from human speech is a challenging task
that, however, is crucial for understanding and modelling speech
production as well as developing speech signal processing algorithms.
The possible approaches can be divided into \emph{direct} and
\emph{indirect methods}. Direct methods concern measurements carried
out on test subjects either by audio recordings, acquisition of
pressure, flow velocity, or even electrical signals (such as takes
place in electroglottography), or using different methods of medical
imaging during speech. Indirect methods concern simulations using
computational models (such as described in \cite{A-A-M-M-V:MLBVFVTOPI}
and the references therein) or measurements from \emph{physical
  models}\footnote{Physical models are understood as artefacts or
  replicas of parts of the speech anatomy in the context of this
  article.}. Typically, computational and physical models are created
and evaluated based on data that has first been acquired by direct
methods. The main advantage of indirect methods is the absence of the
human component that leads to experimental restrictions and unwanted
variation in data quality.

The purpose of this article is to describe an experimental
arrangement, its validation, and some experiments on one type of
physical model for vowel production: \emph{acoustic resonators}
corresponding to vocal tract (VT) configurations during prolonged
vowel utterance. The anatomic geometry for such resonators has been
imaged by Magnetic Resonance Imaging (MRI) with simultaneous speech
recordings as described in
\cite{A-A-H-J-K-K-L-M-M-P-S-V:LSDASMRIS,K-M-O:PPSRDMRI}. The MRI voxel
data has been processed to surface models as explained in
\cite{O-M:ASUAMRIVTGE} and then printed in ABS plastic by Rapid
Prototyping as explained below in Section~\ref{ProcessingSubSec}. In
itself, the idea of using 3D printed VT models in speech research is
by no means new: see, e.g.,
\cite{T-K-E-S-W:EEIFPSKGFTC,E-S-W:NIMARVTDP,T-M-K:AAVTDVPFDTDM}.

\begin{figure}[t]
\begin{center}
\includegraphics[width=0.21\textwidth]{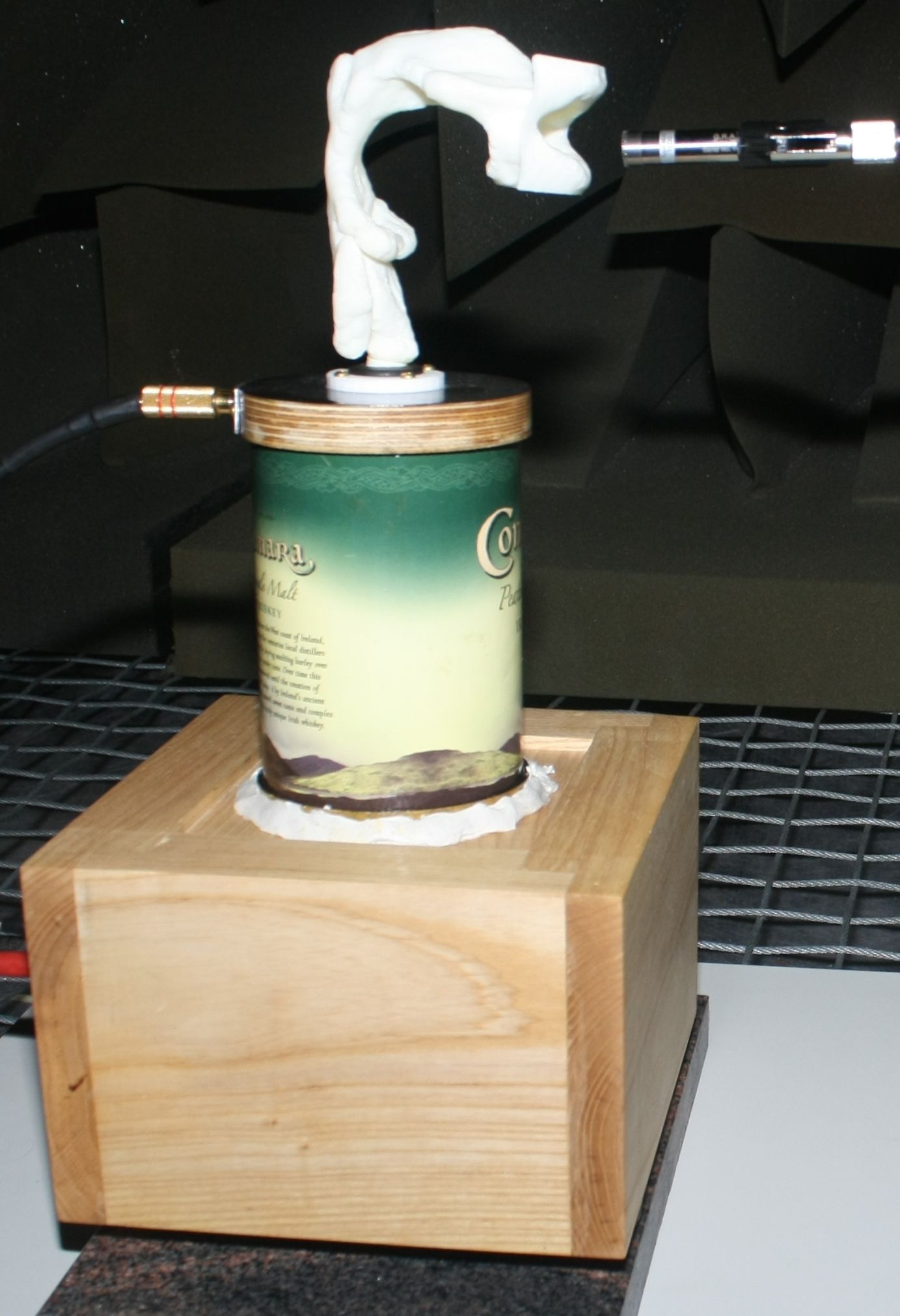}\hspace{0.31cm}
\includegraphics[width=0.43\textwidth]{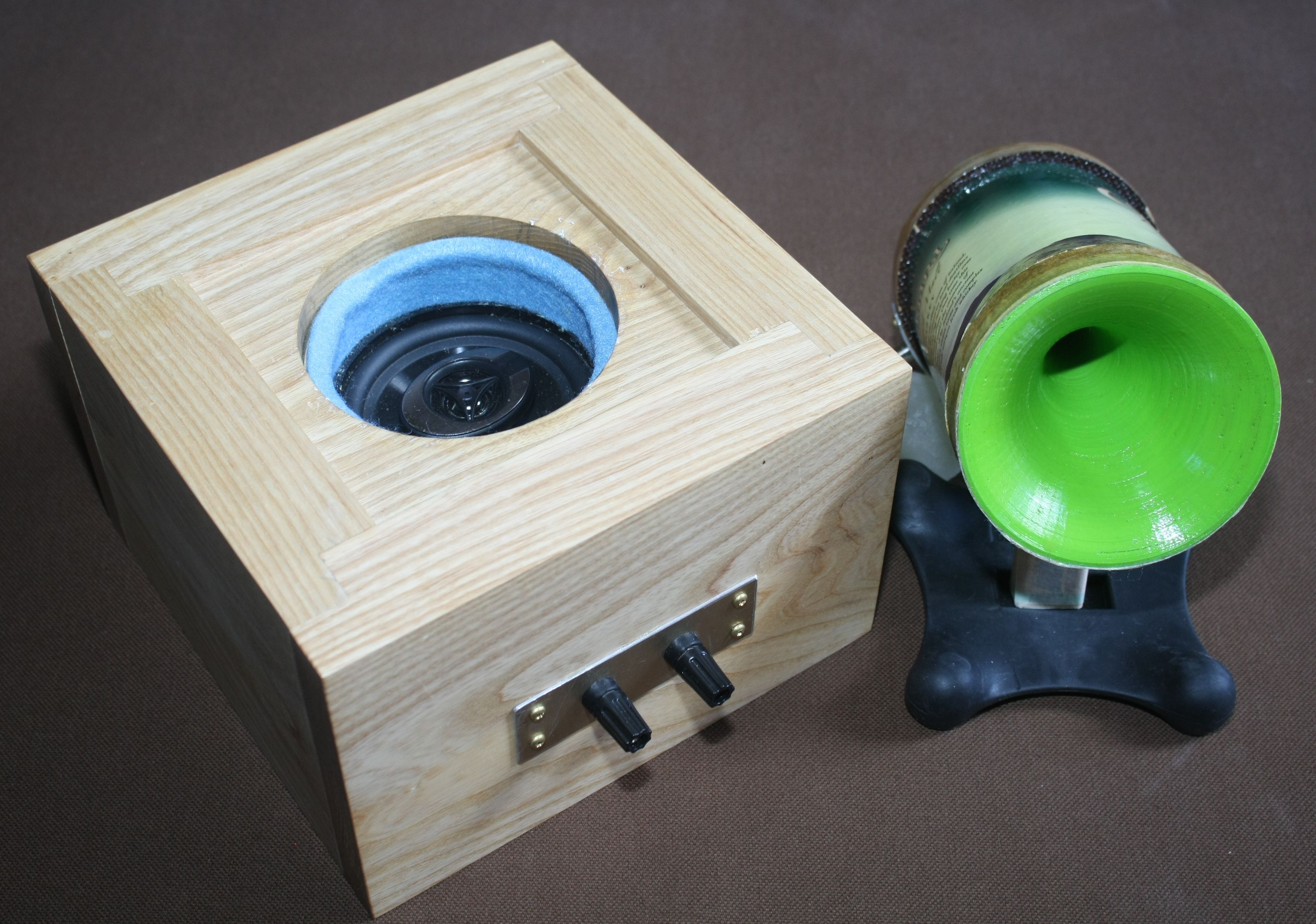}\hspace{0.31cm}
\includegraphics[width=0.29\textwidth]{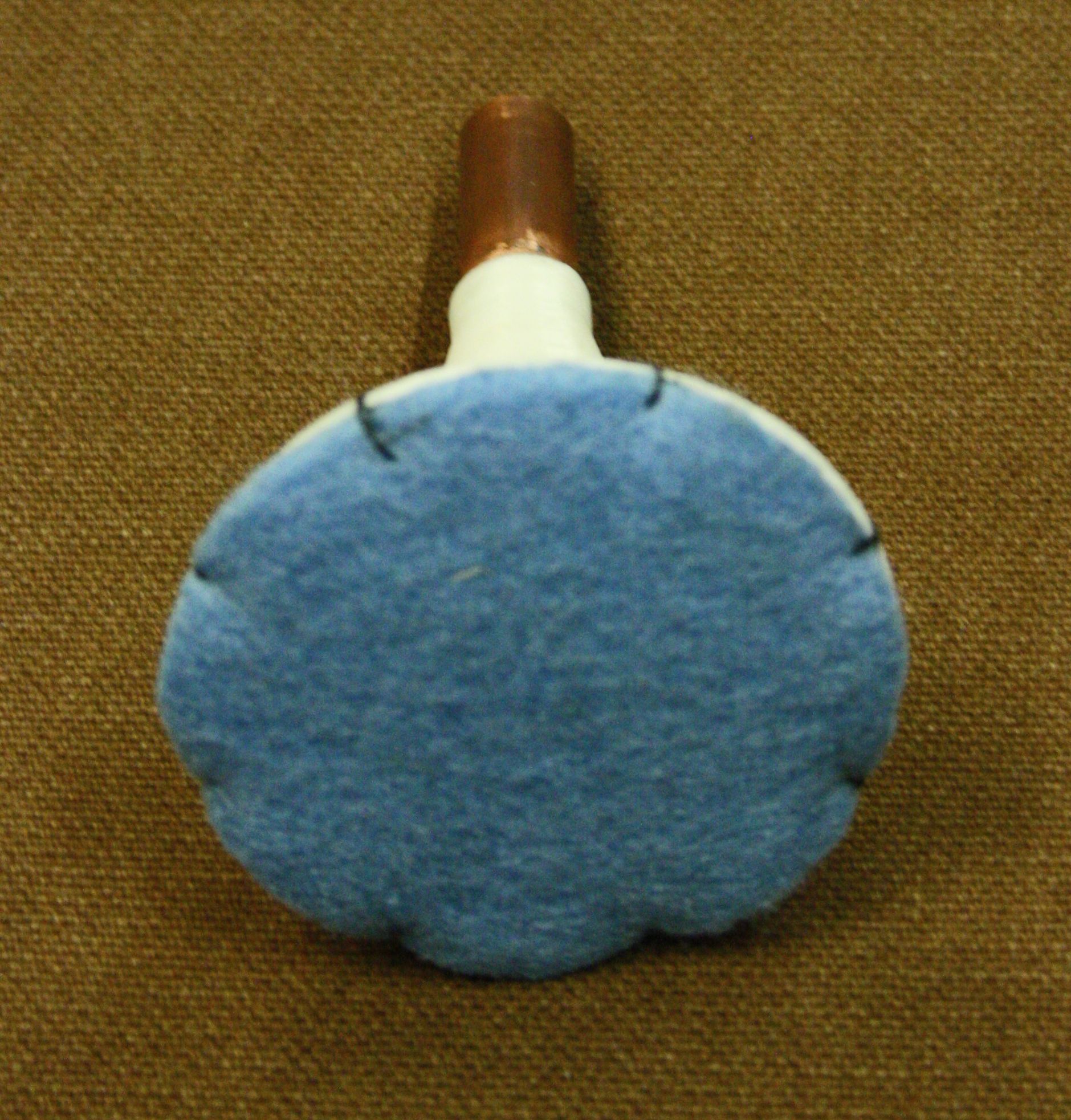}
\end{center}
  \caption{\label{TractrixHornFig} Left: Measurement arrangement for
    the frequency response of vowel [\textipa{\textscripta}] from a 3D
    printed VT geometry. Middle: The tractrix horn and the loudspeaker
    unit assembly separated.  Right: The dummy load used for
    calibration measurements as explained in
    Section~\ref{CalibrationSec}.  }
\end{figure}

Just creating physical models of the VT is not enough for model
experiments: also a suitable acoustic signal source is required with
custom instrumentation and software associated to it. As these
experiments involve a niche area in speech research, directly
applicable commercial solutions do not exists and constructing a
custom measurement suite looks an attractive option. Thus, we propose
an \emph{acoustic glottal source} design shown in
Fig.~\ref{TractrixHornFig} that resembles the loudspeaker-horn
constructions shown in \cite[Fig.~1]{T-K-E-S-W:EEIFPSKGFTC},
\cite[Fig.~3]{E-S-W:NIMARVTDP}, \cite[Fig.~2a]{Wolfe:AIS:2000} . 

All such source/horn constructions can
be regarded as variants of \emph{compression drivers} used as high
impedance sources for horn loudspeakers. Unfortunately, most
commercially available compression drivers are designed for
frequencies over $500 \, \mathrm{Hz}$ whereas a construction based on
a loudspeaker unit can easily be scaled down to lower frequencies
required in speech research.  We point out that high quality acoustic
measurements on VT physical models can be carried out using a
measurement arrangement not based an impedance matching horn or a
compression driver of some other kind; see
\cite[Fig.~3]{T-M-K:AAVTDVPFDTDM} where the sound pressure is fed into
the model through the mouth opening, and the measurements are carried
out using a microphone at the vocal folds position.  However, excitation
from the glottal position is desirable because the face and the
exterior space acoustics are issues as well.

The general principle of operation of the acoustic glottal source is
fairly simple. The source consists of a loudspeaker unit and an
impedance matching horn as shown disassembled in
Fig.~\ref{TractrixHornFig} (middle panel). The purpose of the horn is
to concentrate the acoustic power from the low-impedance loudspeaker
to an opening of diameter $6\, \mathrm{mm}$, the high-impedance output
of the source.  There is, however, a number of conflicting design
objectives that need be taken into account in a satisfactory way. For
example, the instrument should not be impractically large, and it
should be usable for acoustic measurements of physical models of human
VTs in the frequency range of interest, specified as $80
\ldots 7350 \, \mathrm{Hz}$ in this article.  To achieve these goals
in a meaningful manner, we use a design methodology involving
\textrm{(i)} heuristic reasoning based on mathematical acoustics,
together with \textrm{(ii)} numerical acoustics modelling of the main
components and their interactions. Numerical modelling of all details
is not necessary for a successful outcome. Optimising the source
performance using only the method of trial and error and extensive
laboratory measurements would be overly time consuming as well.

The design and construction process was incremental, and it consisted
of the following steps that were repeated when necessary:
\begin{enumerate}
\item[(i)] Choice of the acoustic design and the main components,
  based on general principles of acoustics, horn design, and
  feasibility,
\item[(ii)] Finite element (FEM) based modelling of the horn acoustics
  to check overall validity of the approach, to detect and then
  correct the expected problems in construction,
\item[(iii)] the construction of the horn and the loudspeaker
  assembly together with the required instrumentation,
\item[(iv)] a cycle of measurements and modifications, such as
  placement of acoustically soft material and silicone sealings in
  various parts based on, e.g., the FEM modelling,
\item[(v)] development of MATLAB software for producing properly
  weighted measurement signals for sweep experiments that compensate
  most of the remaining nonidealities, and
\item[(vi)] development of MATLAB software for reproducing the
  Liljencrants--Fant (LF) glottal waveform excitation at the glottal
  position of the physical models.
\end{enumerate}

Finally, the source is used for measuring the frequency responses of
physical models of VT during the utterance of Finnish vowels
[\textipa{\textscripta, i, u}], obtained from a 26-year-old male (in
fact, one of the authors of this article).
%  using MRI as explained in \cite{A-A-H-J-K-K-L-M-M-P-S-V:LSDASMRIS}.  
The measured amplitude frequency responses are compared with the
spectral envelope data from vowel samples shown in
Fig.~\ref{VTResponseFig}, recorded in anechoic chamber from the same
test subject. In addition to these responses, vowel signal is produced
by acoustically exciting the physical models by a glottal pulse
waveform of LF type, reconstructed at the output of the source. The
produced signals for vowels [\textipa{\textscripta, i, u}] have good
audible resolution from each other, yet they have the distinct
``robotic'' sound quality that is typical of most synthetically
produced speech.

Resonant frequencies extracted from the measured frequency responses
are used for development and validation of acoustic and phonation
models such as the one introduced in \cite{A-A-M-M-V:MLBVFVTOPI}. The
synthetic vowel signals are intended for benchmarking Glottal Inverse
Filtering (GIF) algorithms as was done in
\cite{Alku:EstVoiceSrc:2006,Alku:IFReview:2011}. Large amounts of
measurement data are required for these applications which imposes
requirements to the measurement arrangement.

So as to physical dimension of the measured signals, this article
restricts to sound pressure measurements using microphones. If
acoustic impedances are to be measured instead, some form of acoustic
(perturbation) velocity measurement need be carried out. The velocity
measurement can be carried out, e.g., by hot wire anemometers
\cite{Kob:MMV:2002}, impedance heads consisting of several microphones
\cite{Wolfe:EEI:2013}, or even by a single microphone using a
resistive calibration load coupled to a high impedance source
\cite{Singh:AIM:1978}; see \cite[Table~1]{Wolfe:IPM:2006} for various
approaches. In general, carrying out velocity measurements is much
more difficult and expensive that measuring just sound pressure.
Determining pressure-to-pressure -responses of VT physical models is,
however, sufficient for the purposes of this article since
(\textrm{i}) resonant frequencies can be determined from pressures,
and (\textrm{ii}) the GIF algorithm can be configured to run on
pressure data.

\section{\label{BackgroundSec} Background}

We review relevant aspects from mathematical acoustics, horn design,
signal processing, and MRI data acquisition.

\subsection{Acoustic equations for horns}

Acoustic horns are impedance matching devices that can be described as
surfaces of revolution in a three-dimensional space. Thus, they are
defined by strictly nonnegative continuous functions $r = R(x)$ where
$x \in [0, \ell]$, $\ell > 0$ being the length of the horn, and $r$
denoting the radius of horn at $x$. The end $x = 0$ ($x = \ell$) is
the \emph{input end} (respectively, the \emph{output end}) of the
horn.  It is typical, though not necessary, that the function
$R(\cdot)$ is either increasing or decreasing.

There exists a wide literature on the design of acoustic (tractrix)
horns for loudspeakers; see, e.g.,
\cite{Dinsdale:1974,Edgar:1981,Delgado:2000,U-W-B:OVMAH}.  As a
general rule, the matching impedance at an end of the horn is
inversely proportional to the opening area. For uniform diameter
waveguides, the matching impedance coincides with the characteristic
impedance given by $Z_0 = \rho c/A_0$ where $A_0$ is the
intersectional area. The constant $c$ denotes the speed of sound and
$\rho$ is the density of the medium.

To describe the acoustics of an air column in a cavity such as a horn,
we use two (partial) differential equations.  The three dimensional
acoustics is described by the lossless Helmholtz equation in terms of
the velocity potential
\begin{equation} \label{HelmHoltzEq}
  \lambda^2 \phi_\lambda = c^2 \Delta \phi_\lambda \text{ on } \Omega  \quad
   \text{ and } \quad 
  \frac{\partial \phi_\lambda}{\partial \nu}({\bf r}) = 0 \text{ on }
  \partial \Omega \setminus \Gamma_0
\end{equation}
where the acoustic domain is denoted by $\Omega \subset \R^3$ with
boundary $\partial \Omega$.  A part of the boundary, denoted by
$\Gamma_0$, is singled out as an interface to the exterior space. In
horn designs of Section~\ref{HelmholtzCavitySec}, the interface
$\Gamma_0$ is the opening at the narrow output end of the horn. In
Section~\ref{CompValSec}, the symbol $\Gamma_0$ denotes a spherical
interface around the mouth opening. For now, we use the Dirichlet
boundary condition on $\Gamma_0$
\begin{equation} \label{DirichletBndry}
  \phi_\lambda({\bf r}) = 0 \text{ on } \Gamma_0.
\end{equation}
Eqs.~\eqref{HelmHoltzEq}-- \eqref{DirichletBndry} have a countably
infinite number of solutions $(\lambda_j, \phi_j) = (\lambda,
\phi_\lambda) \in \C \times H^1(\Omega) \setminus \{ 0\}$ for $j = 1,
2, \ldots$, and each of the solutions is associated to a
\emph{Helmholtz resonant frequency} $f_j$ of $\Omega$ by $f_j =
\mathrm{Im}{\lambda_j}/2 \pi$.

In addition to acoustic resonances, the acoustic transmission
impedance of the source is important. Because it is more practical to
deal with scalar impedances, we use the lossless Webster's resonance
model for defining it, again in terms of Webster's velocity
potential. It is given for any $s \in \C$ by
\begin{equation} \label{WebsterModel}
\begin{aligned}
  s^2 \psi_s & = \frac{c^2}{A(x)} \frac{\partial}{\partial x} \left (A(x) \frac{\partial \psi_s}{\partial x} \right ) \text{ on } [0,\ell], \\
  - A(0) \frac{\partial \psi_s}{\partial x}(0) & = \hat i(s), \text{ and } R_L A(\ell)
  \frac{\partial \psi_s}{\partial x}(\ell) = \rho s \phi_s(\ell)
\end{aligned}
\end{equation}
where $A(x) = \pi R(x)^2$ is the intersectional area of the horn,
$\rho$ is the density of air, and $R_L \geq 0$ is the termination
resistance at the output end $x = \ell$ \footnote{Because the external
  termination resistance $R_L$ is the only loss term in
  Eq.~\eqref{WebsterModel}, we call the model lossless.}. Again, the
frequencies and Laplace transform domain $s$ variables are related by
$f = \mathrm{Im} s/ 2\pi$. The function $\hat i(s)$ is the Laplace
transform of the (perturbation) volume velocity used to drive the
horn, and the output is similarly given as the Laplace transform of
the sound pressure given by $\hat p(s) = \rho s \phi_s(\ell)$. Now,
the transmission impedance of the horn, terminated to the resistance
$R_L > 0$, is given by
\begin{equation} \label{TransmissionImpedance}
  Z_{R_L}(s) = \hat p(s)/\hat i(s) \text{ for all } s \in \C_+.
\end{equation}
Note that when solving Eq.~\eqref{WebsterModel} for a fixed $s$, we may
by linearity choose $\hat i(s) = 1$ when plainly $Z_{R_L}(s) = \rho s
\phi_s(\ell)$.  Further, as an impedance of a passive system, the
transmission impedance satisfies the positive real condition
\begin{equation}
 \mathop{Re} {Z_{R_L}(s)} \geq 0 \text{ for all } s \in \cplus := \{ s
 \in \C: \mathop{Re}{s} > 0 \}.
\end{equation}

\begin{figure}[t]
\begin{center}
\includegraphics[scale=0.35]{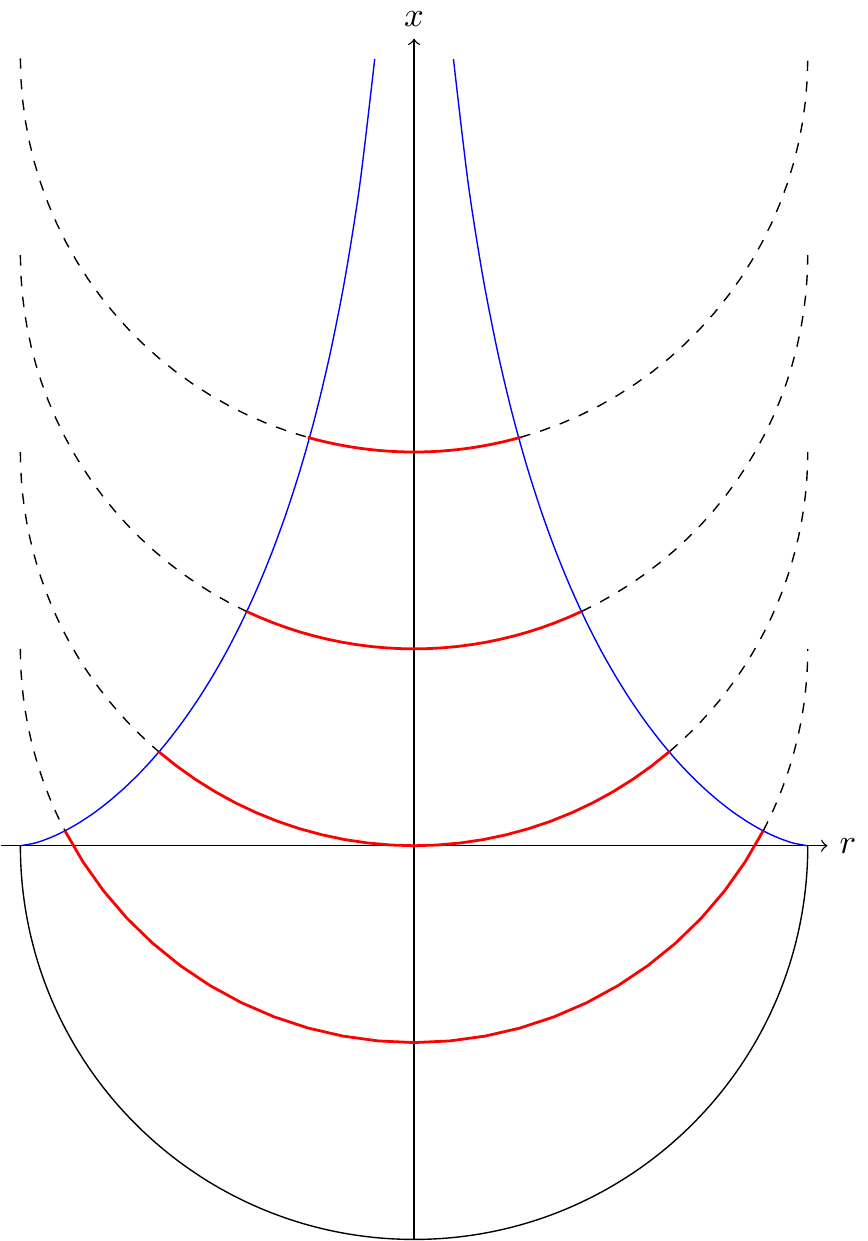} \hspace{0.5cm}
\includegraphics[width=0.20\textwidth]{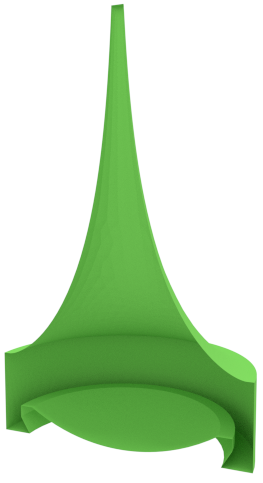} \hspace{0.5cm}
\includegraphics[width=0.33\textwidth]{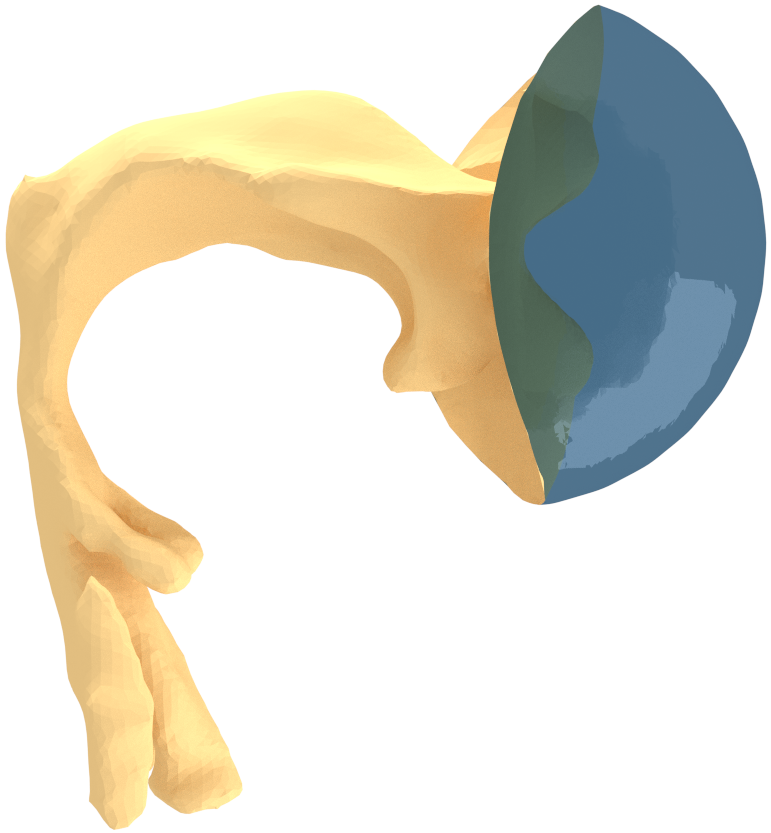} 
\end{center}
  \caption{\label{DesignOfSourceFig} Left: Wave propagation in a
    tractrix horn. The spherical wave front progressing along the
    centreline meets the horn surface at right angles.  Middle: A 3D
    illustration of the impedance matching cavity within the
    source. Right: The geometry of the VT corresponding to vowel
    [\textipa{\textscripta}], equipped with a spherical boundary
    condition interface at the mouth opening.}
\end{figure}

%  The    mid-section of the horn (with dimensions \DisplayNote{JM: missing
%      from the picture}) used for as the impedance matching cavity of
%    the sound source.

\subsection{\label{SupressionSubSec} Suppression of transversal modes in horns}

By transversal modes we refer to the resonant standing wave patterns
in a horn where significant pressure variation is perpendicular to the
horn axis, as opposed to purely longitudinal modes. The purpose of
this section is to argue why transversal modes in horn geometries are
undesirable from the point of view of the this article.

As a well-known special case, consider a wave\-guide of length $\ell$
that has a constant diameter, i.e., $A(x) = A_0$. Then the
transmission impedance given by
Eqs.~\eqref{WebsterModel}--\eqref{TransmissionImpedance} can be given
the explicit formula
\begin{equation} \label{TransmissionLineImpedance}
  Z_{R_L}(s) = \frac{Z_0 R_L}{Z_0 \cosh{\frac{s \ell}{c}} + R_L \sinh{\frac{s \ell}{c}}}
\end{equation}
where $Z_0 := \rho c/A_0$ is called \emph{characteristic
  impedance}. Because both $\cosh$ and $\sinh$ are entire functions,
it is impossible to have $Z_{R_L}(s) = 0$ for any $s \in \C$. If the
termination resistance $R_L$ equals the characteristic impedance of
the wave\-guide, the wave\-guide becomes nonresonant, and we get the
pure delay $Z_{R_L}(s) = Z_0 e^{- s \ell / c}$ of duration $T = \ell /
c$ as expected.

It can be shown by analysing the Webster's model that the transmission
impedance $Z_{R_L}(s)$ given by Eq.~\eqref{TransmissionImpedance} has
no zeroes for $s \in \C$; i.e., it is an all-pole transmission impedance
for any finite value of termination resistance
$R_L > 0$.\footnote{This follows from Holmgren's uniqueness theorem
  for real analytic area functions $A(\cdot)$.}  The salient, desirable
feature of any all-pole impedance is that also the admittance
$A_{R_L}(s) := Z_{R_L}(s)^{-1}$ is analytic and even
$\mathop{Re}{A_{R_L}(s)} > 0$ in $s \in \cplus$. This makes it easy to
precompensate the lack of flatness in the frequency response of
$Z_{R_L}(s)$ by a causal, passive, rational filter whose transfer
function approximates $A_{R_L}(s)$.

On the other hand, it has been shown in
\cite[Theorem~5.1]{L-M:PEEWEWP} that the time-dependent Webster's
model describes accurately the transversal averages of a 3D wavefront
in an acoustic wave\-guide if the wavefront itself is constant on the
transversal sections of the wave\-guide interior. Conversely,
Webster's equation models only the longitudinal dynamics of the
wave\-guide acoustics by its very definition as can be understood
from, e.g., \cite{L-M:WECAD}. If the transversal modes in a
wave\-guide have been significantly excited, then Webster's equation
becomes a poor approximation, and all hopes of regarding the measured
transmission impedance of the wave\-guide as an all-pole SISO system
are lost. A more intuitive way of seeing why transversal acoustic
modes are expected to introduce zeroes to $Z_{R_L}(\cdot)$ is by
reasoning by analogy with Helmholtz resonators: the resonant side
branches of the wave\-guide (eliciting transversal modes at desired
frequencies) can be used to eliminate frequencies from response.

We have now connected, via Webster's horn model, the appearance of
transversal modes in a horn to zeroes of the transmission impedance
$Z_{R_L}(\cdot)$.  Because these zeroes are undesirable features in
good horn designs, we need to identify and suppress the transversal
modes as well as is feasible.

\subsection{Minimisation of transmission loss}
\label{TLMinSec}

When a horn is excited from its input end, some of the excitation
energy is reflected back to the source with some delays.  For horns of
finite length $\ell$, there are two kinds of backward
reflections. Firstly, the geometry of the horn may cause distributed
backward reflections over the the length of the horn.  Secondly, there
may be backward reflections at the output end of the horn, depending
on the acoustic impedance seen by the horn at the termination point $x
= \ell$ in Eq.~\eqref{WebsterModel}. We next consider only the
backward reflections of the first kind since only they can be affected
by the horn design.

Because the acoustics of the horn described by
Eqs.~\eqref{HelmHoltzEq}---\eqref{WebsterModel} is internally
lossless, minimising the TL amounts to minimising the backward
reflections that take place inside the horn. This is a classical shape
optimisation problem in designing acoustical horns, and modern
approaches are based on numerical topology optimisation techniques as
presented in, e.g., \cite{U-W-B:OVMAH,Y-W-B:LOTSHMIFFDPAH} where also
other design objectives (typical of loudspeaker horn design) are
typically taken into account.

We take another approach, and use analytic geometry and physical
simplifications of wave propagation for designing the function $r =
R(x)$ on $[0, \ell]$ following Paul~G.~A.~H.~Voigt who proposed a
family of tractrix horns in his patent ``Improvements in Horns for
Acoustic Instruments'' in 1926, see \cite{PV:IHAI}. His invention was
to use the surface of revolution of tractrix curve given by
\begin{equation} \label{TractrixEq}
  x = a \ln{\frac{a + \sqrt{a^2 - r^2}}{r}} - \sqrt{a^2 - r^2}, \quad r \in [0, a]
\end{equation}
where $a > 0$ is a parameter specifying the radius of the wide (input)
end. Obviously, Eq.~\eqref{TractrixEq} defines a decreasing function
$x \mapsto R(x) = r$ mapping $R:[0,\infty) \to (0,a]$ with $R(0) = a$
and $\lim_{x \to \infty}{R(x)} = 0$ which defines the \emph{tractrix
  horn}. The required finite length $\ell > 0$ of the horn is solved
from $R(\ell) = b$ where $0 < b < a$ is the required radius of the
(narrow) output end.

The tractrix horn is known as the \emph{pseudosphere} of constant
negative Gaussian curvature in differential geometry. That it acts as
a spherical wave horn is based on Huyghens principle and a geometric
property of Eq.~\eqref{TractrixEq}. More precisely, it can be seen
from Fig.~\ref{TractrixHornFig} (left panel) that a spherical wave
front of curvature radius $a$, propagating along the centreline of the
horn, meets the tractrix horn surfaces always in right
angles. Disregarding, e.g., the viscosity effects in the boundary
layer at the horn surface, the right angle property is expected to
produce minimal backward reflections for spherical waves similarly as
a planar wavefront would behave in a constant diameter wave\-guide far
away from wave\-guide walls.

\subsection{\label{DeConvSec} Regularised deconvolution}

A desired sound waveform target pattern will be reconstructed at the
source output by compensating the source dynamics in
Section~\ref{ImpulseSubSec}.  Our approach is based on the idea of
\emph{constrained least squares filtering} used in digital image
processing \cite{Hunt:DLS:1972,Phillips:TNS:1962}.

Suppose that a linear, time-invariant system has the real-valued
impulse response $h(t) = h_0(t) + h_e(t)$ that is expected to contain
some measurement error $h_e(t)$. When the input signal $u = u(t)$ is
fed to the system, the measured output is obtained from
\begin{equation} \label{ConvolutionEq}
y(t) = (h_0*u)(t) + v(t) \quad \text{ with } \quad v = h_e * u + w
\quad \text{ for } \quad t \in [0, T].
\end{equation}
As usual, the convolution is defined by $(h_0*u)(t) = \int_{-\infty}^t
{h_0(t - \tau) u(\tau) \, d \tau}$, and our task is to estimate $u$
from Eq.~\eqref{ConvolutionEq} given $y$ and some incomplete
information about the output noise $v$.  We assume $u, v \in L^2(0,T)$
and that $h_0$ is a continuous function.  We define the noise level
parameter by $\epsilon = \Vert v \Vert_{L^2(0,T)} / \Vert y \Vert_{L^2(0,T)}$ and require that $0
< \epsilon < 1$ holds.

Unfortunately, Eq.~\eqref{ConvolutionEq} is not typically solvable for
smooth $y$ since the noise $v$ is not generally even continuous
whereas the convolution operator $h_0*$ is smoothing. Instead of
solving Eq.~\eqref{ConvolutionEq}, we solve an estimate $\v u$ for $u$
from the regularised version of Eq.~\eqref{ConvolutionEq}, given for
$y \in L^2(0,T)$ by
\begin{equation} \label{RegularisedEq}
  \begin{aligned}
    \mathrm{Arg \, min} & \left ( \kappa \Vert \v u \Vert_{L^2(0,T)}^2 + \Vert
    \v u'' \Vert_{L^2(0,T)}^2 \right ) \\
    & \text{ with the constraint } \Vert
    y - h_0* \v u \Vert_{L^2(0,T)} = \epsilon \Vert y \Vert_{L^2(0,T)}.
  \end{aligned}
\end{equation}
Here $T> 0$ is the sample length, $\kappa > 0$ is a regularisation
parameter, and $\epsilon$ is the noise level introduced above in the
view of $v$ in Eq.~\eqref{ConvolutionEq}. Obviously, it is not
generally possible to choose $\epsilon = 0$ in
Eq.~\eqref{RegularisedEq} without rendering $y = h_0* \v u$
insolvable in $L^2(0,T)$.

Using Lagrange multipliers, the Lagrangian function takes the form
\begin{equation*}
  L_\epsilon(\v u, \mu) = \kappa \Vert \v u \Vert_{L^2(0,T)}^2 
  + \Vert \v u'' \Vert^2_{L^2(0,T)}  
  -\mu \left (\Vert y - h_0* \v u \Vert_{L^2(0,T)}^2 - \epsilon^2 \Vert y \Vert^2_{L^2(0,T)} \right ).
\end{equation*}
Using the variation $\tilde{u}_\eta = \v u+\eta w$ with $\eta
\in \R$, we get
\begin{equation*}
\begin{aligned}
  & \frac{d}{d\eta} L_\epsilon (\tilde{u}_{\eta}, \mu)
  \bigg|_{\eta=0} \\
= &  2\mathrm{Re} \left( \kappa \langle  w , \v u \rangle_{L^2(0,T)} + \langle  w'' , \v u'' \rangle_{L^2(0,T)}
     -\mu \langle h_0* w, y-h_0 *\v u  \rangle_{L^2(0,T)}
  \right) = 0\,
\end{aligned}
\end{equation*}
for all test functions $w \in \mathcal{D}([0,T])$. Thus
\begin{equation*}
  \kappa \langle w, \v u \rangle + \langle w'', \v u'' \rangle - \mu \langle  h_0*w, y-h_0*\v u \rangle=0
\end{equation*}
which, after partial integration and adjoining the convolution
operator $h_0*$, gives
\begin{equation*}
\kappa \v u + \v u^{(4)} -\mu (h_0*)^* \left( y- h_0*\v u \right)
  = 0,
\end{equation*}
leading to the normal equation
\begin{equation} \label{NormalEq}
\v u = \left[\gamma \left ( \kappa + \frac{d^4}{dt^4} \right) +
  (h_0*)^*(h_0*) \right]^{-1}(h_0*)^*y
\end{equation}
together with the constraint $\Vert y - h_0* \v u \Vert_{L^2(0,T)} =
\epsilon \Vert y \Vert_{L^2(0,T)}$ where $\gamma= \gamma(y,\epsilon) \in \R$ satisfies $\gamma
= 1/ \mu$ (a constant independent of $t$). By a direct computation
using commutativity, we get for the residual
\begin{equation} \label{ResidualEq}
  v_{\kappa, \mu} = y - h_0 * \v u = \left (\kappa + \frac{d^4}{dt^4}
  + \mu (h_0*)^*(h_0*) \right )^{-1} \left (\kappa y +  y^{(4)} \right ).
\end{equation}
Because $\gamma, \kappa > 0$, the inverses in
Eqs.~\eqref{NormalEq}--\eqref{ResidualEq} exist by positivity of the
operators.

So, the possible noise components $v$ in Eq.~\eqref{ConvolutionEq},
consistent with Eq.~\eqref{NormalEq}, are the two parameter family $v
= v_{\kappa, \mu}$ given in Eq.~\eqref{ResidualEq} where $\kappa, \mu
> 0$. For each $\kappa$, we have
\begin{equation*}
  \Vert v_{\kappa, 0} \Vert_{L^2(0,T)} 
  = \Vert y \Vert_{L^2(0,T)} 
  \text{ and } \lim_{\mu \to \infty} { \Vert v_{\kappa, \mu} \Vert_{L^2(0,T)}} = 0.
\end{equation*}
By continuity and the inequality $0 < \epsilon < 1$, there exists a
$\mu_0 = \mu_0(\epsilon, \kappa)$ such that $\Vert v_{\kappa, \mu_0}
\Vert_{L^2(0,T)} = \epsilon \Vert y \Vert_{L^2(0,T)}$ as required.  We
conclude that $\v u$ given by Eq.~\eqref{NormalEq} with $\gamma =
1/\mu_0$ is a solution of the optimisation problem
\eqref{RegularisedEq}, and, hence, the regularised solution of
Eq.~\eqref{ConvolutionEq} depending on parameters $\epsilon, \kappa >
0$. In practice, the values of these regularising parameters must be
chosen based on the original problem data $y$ and $v$.

In frequency plane, Eqs.~\eqref{NormalEq}--\eqref{ResidualEq} take the
form
\begin{equation*}
  \hat u(\xi) = \frac{\overline{H(i \xi)} \hat y (\xi)}
       {\gamma \left ( \kappa +   \xi^4 \right) + \abs{H(i \xi)}^2} 
\end{equation*}
where $H(s) = \int_0^\infty {e^{-st}h_0(t) \, dt} $ is the transfer
function corresponding to $h_0(t)$ and
\begin{equation}\label{remainderEq}
  \hat  v_{\kappa,\mu}(\xi) = G_{\kappa,\mu}(i \xi) \hat y (\xi) \quad \text{ where } \quad
  G_{\kappa,\mu}(s) = \left (1 + \frac{\mu \abs{H(s)}^2}{ \kappa + s^4} \right )^{-1}.
\end{equation}
Note that $\abs{G_{\kappa,\mu}(i \xi)} < 1$, and the last equation
indicates that the high frequency components of $y$ and
$v_{\kappa,\mu}$ are essentially identical. By Parseval's identity,
the value of $\gamma = 1/\mu_0$ is solved from $\frac{1}{2
  \pi}\int_{-\infty}^\infty {\abs{\hat  v_{\kappa,\mu}(\xi)}^2
  \, d \xi } = \epsilon^2  \Vert y \Vert^2_{L^2(0,T)} $.

\subsection{\label{ProcessingSubSec} Processing of VT anatomic data and sound}

Three-dimensional anatomic data of the VT is used for
computational validations of the sound source as well as for carrying
out measurements using physical models.

VT anatomic geometries were obtained from a (then)
26-year-old male (in fact, one of the authors of this article) using
3D MRI during the utterance of Finnish vowels [\textipa{\textscripta,
    i, u}] as explained in \cite{A-A-H-J-K-K-L-M-M-P-S-V:LSDASMRIS}. A
speech sample was recorded during the MRI, and it was processed for
formant analysis by the algorithm described in \cite{K-M-O:PPSRDMRI}.
The formant extraction for Section~\ref{CompValSec} was carried out
using Praat \cite{Praat:2016}.  Three of the MR images
corresponding to Finnish quantal vowels [\textipa{\textscripta, i, u}]
were processed into 3D surface models (i.e., STL files) as explained
in \cite{O-M:ASUAMRIVTGE}. A spherical boundary condition interface
was attached at the mouth opening for the geometry corresponding
[\textipa{\textscripta}] for producing the computational geometries
shown in Fig.~\ref{CoupledSystemRes}.

\begin{figure}[t]
\begin{center}
\includegraphics[width=0.28\textwidth]{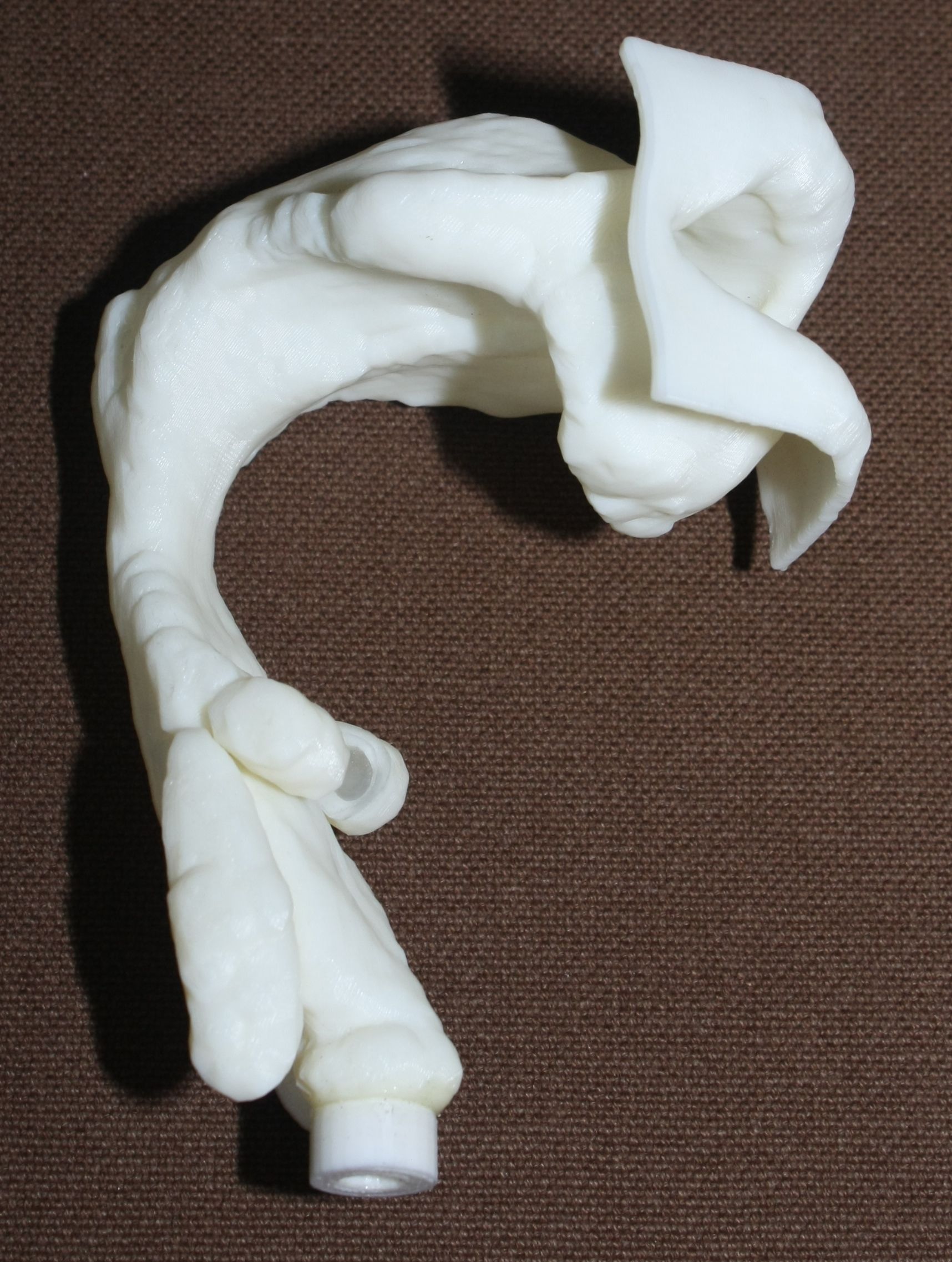}\hspace{0.2cm}
\includegraphics[width=0.277\textwidth]{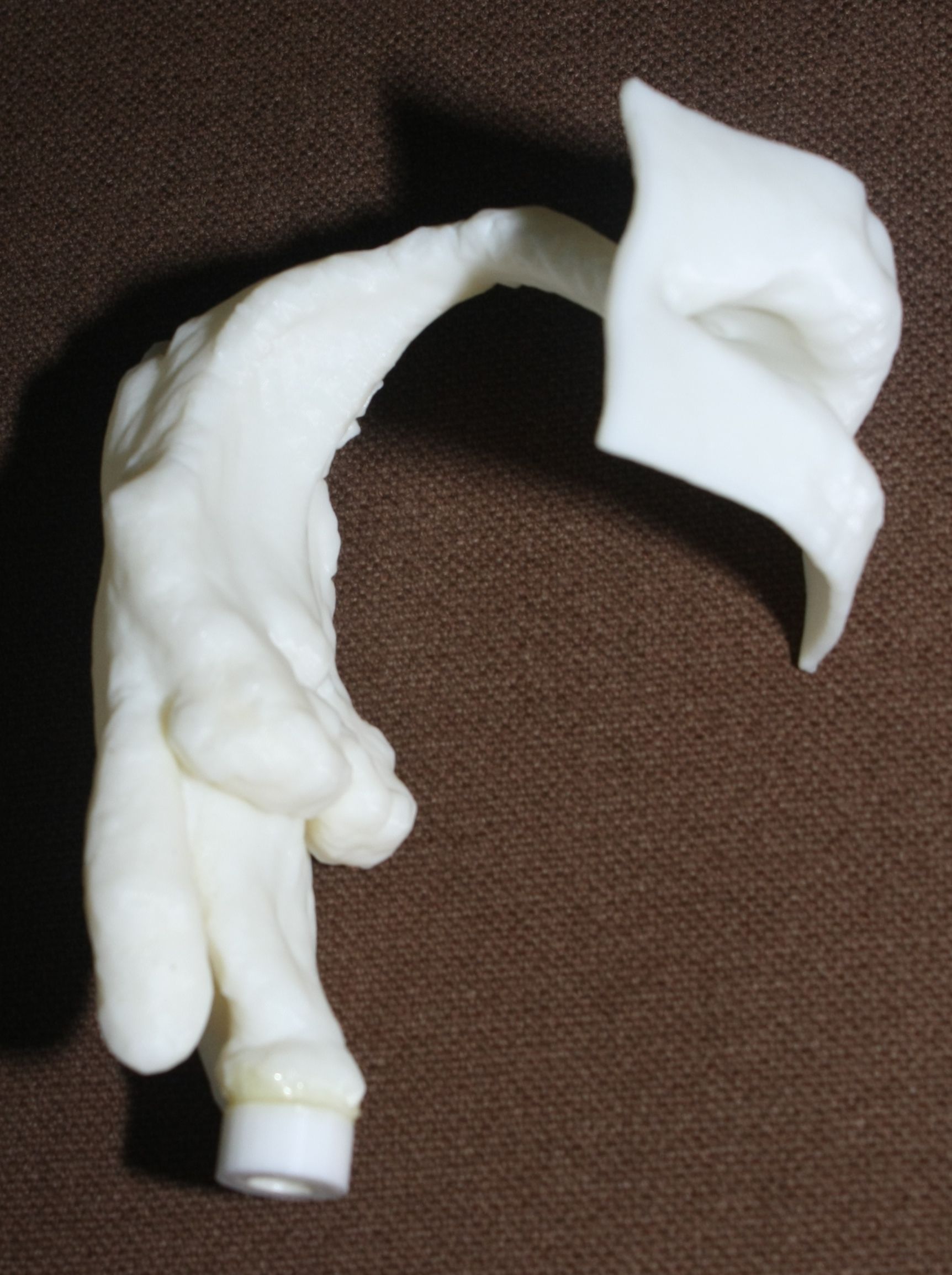}\hspace{0.2cm}
\includegraphics[width=0.345\textwidth]{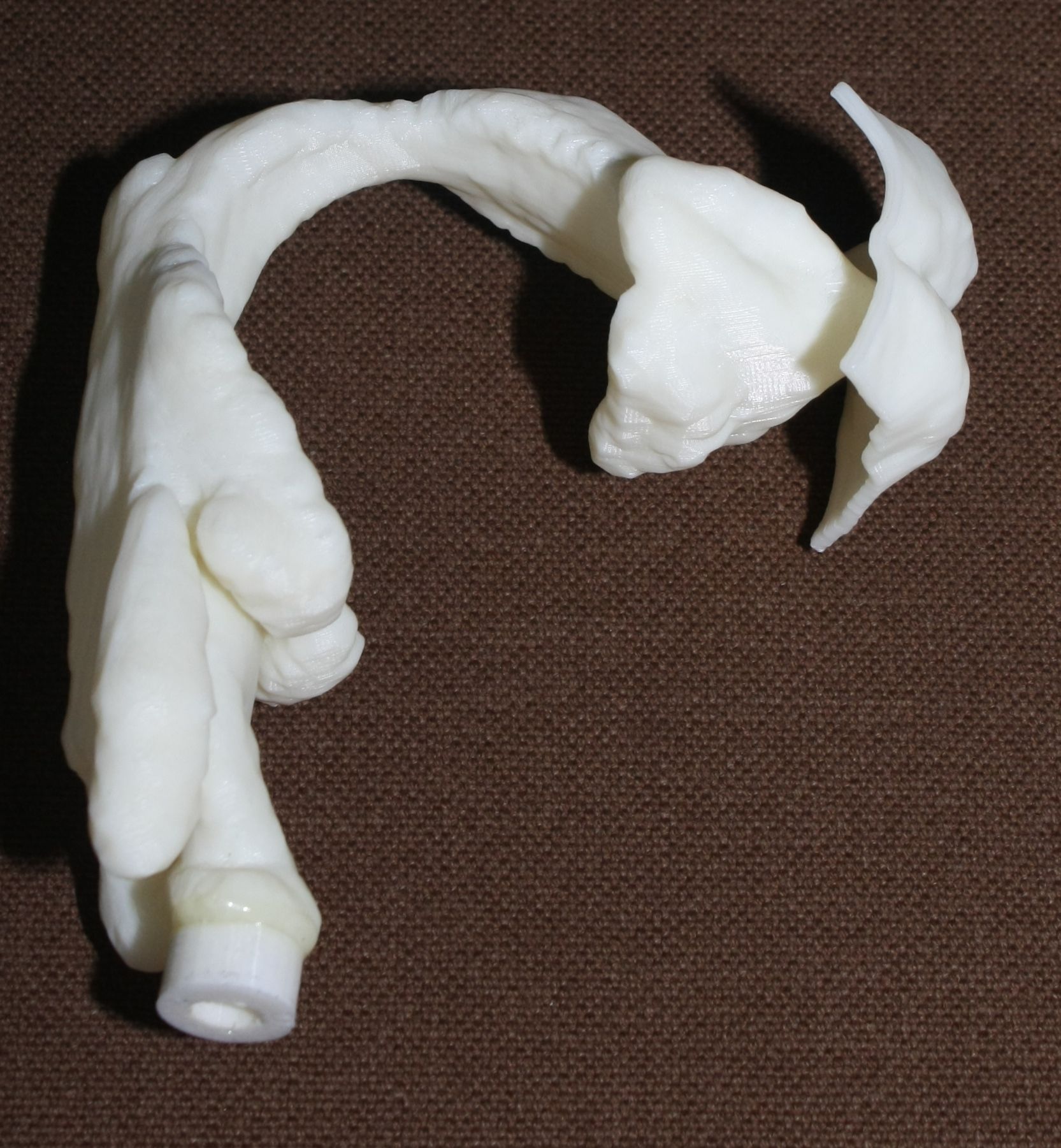} \hspace{0.2cm}
\end{center}
  \caption{\label{VTPrints} Physical VT models of
    articulation geometries corresponding to [\textipa{\textscripta,
        i, u}]. Adaptor sleeves have been glued to the glottis ends
    for coupling to the sound source.}
\end{figure}

Stratasys uPrint SE Plus 3D printer was used to produce physical
models in ABS plastic from the STL files, shown in
Fig.~\ref{VTPrints}.  The printed models are in natural scale with
wall thickness $2 \, \mathrm{mm}$, they extend from the glottal
position to the lips, and they were equipped with an adapter (visible
in Fig.~\ref{VTPrints}) for coupling them to a acoustic sound source
shown in Fig.~\ref{TractrixHornFig} (left panel).

\section{Design and construction}

Based on the considerations of Section~\ref{BackgroundSec}, we
conclude that the following three design objectives are desirable for
achieving a successful design:
%  for an acoustical impedance matching device:
\begin{enumerate}
\item[(i)] \label{Req1} The transmission loss (henceforth, TL) from
  the the input to output should be as low as possible.
\item[(ii)] \label{Req2} There should be no strong transversal
  resonant modes inside the impedance matching cavity of the device.
\item[(iii)] \label{Req3} The frequency response
  $\omega \mapsto \abs{Z_{R_L}(i \omega)}$ of the transmission
  impedance should be as flat as possible for relevant termination
  resistances.
\end{enumerate}
It is difficult --- if not impossible --- to optimise all these
characteristics in the same device.  Fortunately, DSP techniques can
be used to cancel out some undesirable features, and instead of
requirement \textrm{(iii)} it is more practical to pursue a more
modest goal:
\begin{enumerate}
\item[(iii')] \label{Req3weaker} The frequency response  $\omega \mapsto
  \abs{Z_{R_L}(i \omega)}$ should be such that its lack of flatness can be
  accurately precompensated by causal, rational filters.
\end{enumerate}
We next discuss each of these design objectives and their solutions in
the light of Section~\ref{BackgroundSec}.

The tractrix horn geometry was chosen so as to minimise the TL as
explained in Section~\ref{TLMinSec}.  In the design proposed in this
article, we use $a = 50.0 \, \mathrm{mm}$, $b = 2.2 \, \mathrm{mm}$,
and $\ell = 153.0 \, \mathrm{mm}$ as nominal values in
Eq.~\eqref{TractrixEq}. The physical size was decided based on reasons
of practicality and the availability of suitable loudspeaker units.

Contrary to horn loudspeakers or gramophone horns having essentially
point sources at the narrow input end of the horn, the sound source is
now located at the wide end of the horn. Hence, it would be desirable
to generate the acoustic field by a spherical surface source of
curvature radius $a$ whose centrepoint lies at the centre of the
opening of the wider input end. This goal is impossible to precisely
attain using commonly available loudspeaker units, but a reasonable
outcome can be obtained just by placing the loudspeaker (with a
conical diaphragm) at an optimal distance from the tractrix horn as
shown in Fig.~\ref{DesignOfSourceFig} (middle panel). This results in
a design where the \emph{impedance matching cavity} of the source is a
horn as well, consisting of the tractrix horn that has been extended
at its wide end by a cylinder of diameter $2 a = 100.0 \, \mathrm{mm}$
and height $h = 20.0 \, \mathrm{mm}$. Thus, the total longitudinal
dimension of the impedance matching cavity inside the sound source is
$\ell_{tot} = \ell + h = 173.0 \, \mathrm{mm}$ as shown in
Fig.~\ref{DesignOfSourceFig} (middle panel). This dimension
corresponds to the quarter wavelength resonant frequency at $f_{low} =
1648 \, \mathrm{Hz}$, obtained by solving the eigenvalue problem
Eq.~\eqref{HelmHoltzEq} by finite element method (FEM) shown in
Fig.~\ref{HornSystemRes} (left panel). For frequencies much under
$f_{low}$, the impedance matching cavity need not be considered as a
wave\-guide but just as a delay line.

Since the geometry of impedance matching cavity has already been
specified, there is no \emph{geometric} degrees of freedom left for
improving anything. Thus, it is unavoidable to relax design
requirement \textrm{(iii)} in favour of the weaker requirement
\textrm{(iii')}.  As discussed in Section~\ref{SupressionSubSec},
requirement \textrm{(iii')} can, however, be satisfactorily achieved
if overly strong transversal modes of the impedance matching cavity
can be avoided, i.e., the design requirement \textrm{(ii)} is
sufficiently well met.

%% \DisplayNote{JK: After measuring stuff, I would say that our
%%   $R(0)\approx 50mm$ and $R(\ell) \approx 2.5mm$. Due to construction
%%   it is little hard to measure the horn height, but
%%   $\ell \approx 153mm$. According the Tractrix-formula
%%   $R(153) \approx 2.2$. }
%% Here $R(0) = 70 \,
%% \mathrm{mm}$ is the radius of the loudspeaker unit, and the radius
%% $R(\ell) = 3 \, \mathrm{mm}$ is dictated by the intended use of the
%% instrument as a glottal source shown in Fig.~\ref{TractrixHornFig}
%% (right). The length $\ell$ of the horn is determined by the choice of
%% the geometry; i.e., the function $R$.

\subsection{\label{HelmholtzCavitySec} Modal analysis of the impedance matching cavity}

The first step in treating transversal modes of the impedance matching
cavity is to detect and classify them.  Understanding the modal
behaviour helps the optimal placement of attenuating material. For
this purpose, the Helmholtz equation \eqref{HelmHoltzEq} was solved by
FEM in the geometry of the impedance matching cavity, producing
resonances up to $8 \, \mathrm{kHz}$. Some of the modal pressure
distributions are shown in Fig.~\ref{HornSystemRes}.  As explained in
Section~\ref{CompValSec} below, also the acoustic resonances of the VT
geometry shown Fig.~\ref{DesignOfSourceFig} (right panel) were
computed in a similar manner, and their perturbations were evaluated
when coupled to the impedance matching cavity as shown in
Fig.~\ref{CoupledSystemRes}.

The triangulated surface mesh of the impedance matching cavity was
created by generating a profile curve of the tractrix horn in MATLAB,
from which a surface of revolution was created in Comsol where the
cylindrical space and the loudspeaker profile were included.
Similarly, the surface mesh of the VT during phonation of the Finnish
vowel \textipa{[\textscripta]} was extracted from MRI
data~\cite{O-M:ASUAMRIVTGE}. This surface mesh was then attached to
the surface mesh of the spherical interface $\Gamma{_0}$ shown in
Fig.~\ref{DesignOfSourceFig} (right panel). For computations required
in Section~\ref{CompValSec}, the two surface meshes (i.e., the cavity
and the VT) were joined together at the output end of the tractrix
horn and the glottis, respectively. Finally, tetrahedral volume meshes
for FEM computations were generated using GMSH~\cite{gmsh} of all of
the three geometries with details given in Table~\ref{MeshTable}.

\begin{table}[h]
 \centerline{
    \begin{tabular}{|l|c|c|c|c|}
\hline
 & tetrahedrons  &  d.o.f.  \\
\hline
\textbf{Impedance matching cavity}  & 71525  & 15246 \\
\textbf{Cavity joined with VT} & 175946  & 38020 \\
\textbf{VT} & 97847  & 21745 \\
\hline
\end{tabular}}
\caption{\label{MeshTable} The number of tetrahedrons of the three FEM
  meshes used for resonance computations in
  Sections~\ref{HelmholtzCavitySec}~and~\ref{CompValSec}. The degrees
  of freedom indicates the size of resulting system of linear
  equations for eigenvalue computations.  }
\end{table}

The Helmholtz equation \eqref{HelmHoltzEq} with the Dirichlet boundary
condition \eqref{DirichletBndry} at the output interface $\Gamma_0$ is
solved by FEM using piecewise linear elements. In this case, the
problem reduces to a linear eigenvalue problem whose lowest
eigenvalues give the resonant frequencies and modal pressure
distributions of interest.  Some of these are shown in
Fig.~\ref{HornSystemRes}.

\begin{figure}[t]
\begin{center}
\includegraphics[width=0.22\textwidth]{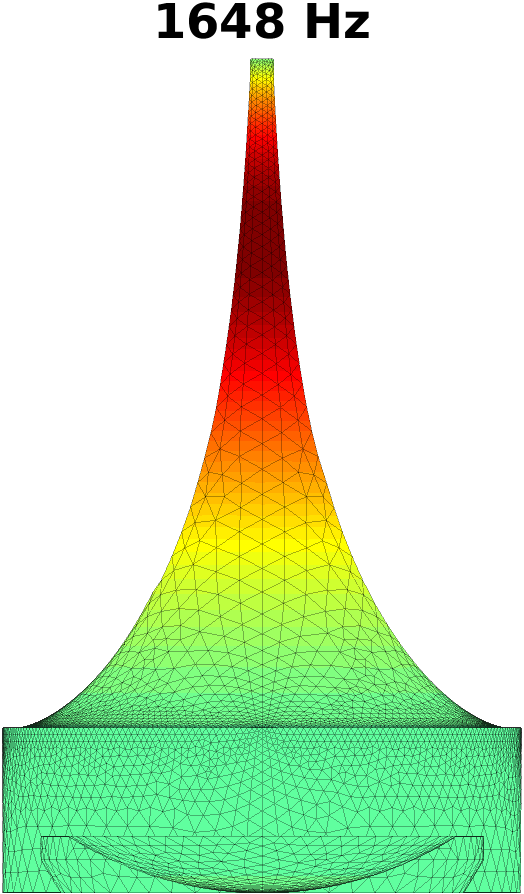}\hspace{0.2cm}
\includegraphics[width=0.22\textwidth]{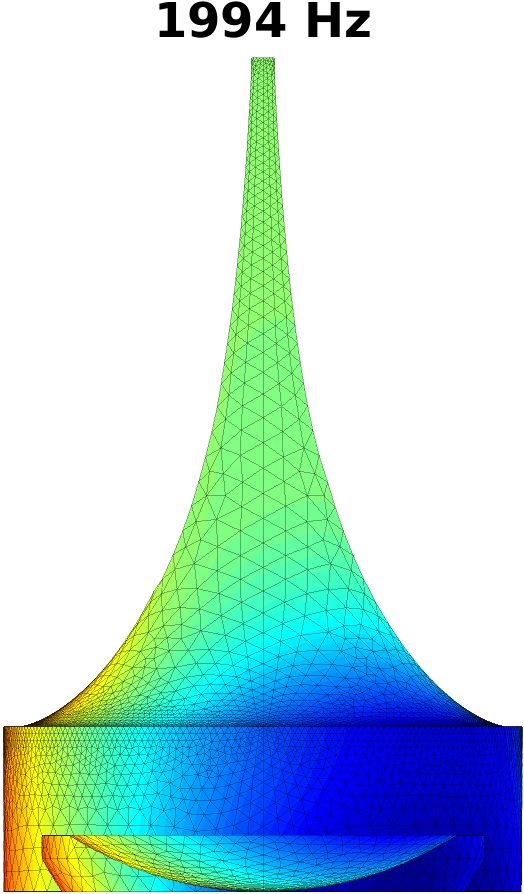}\hspace{0.2cm}
\includegraphics[width=0.22\textwidth]{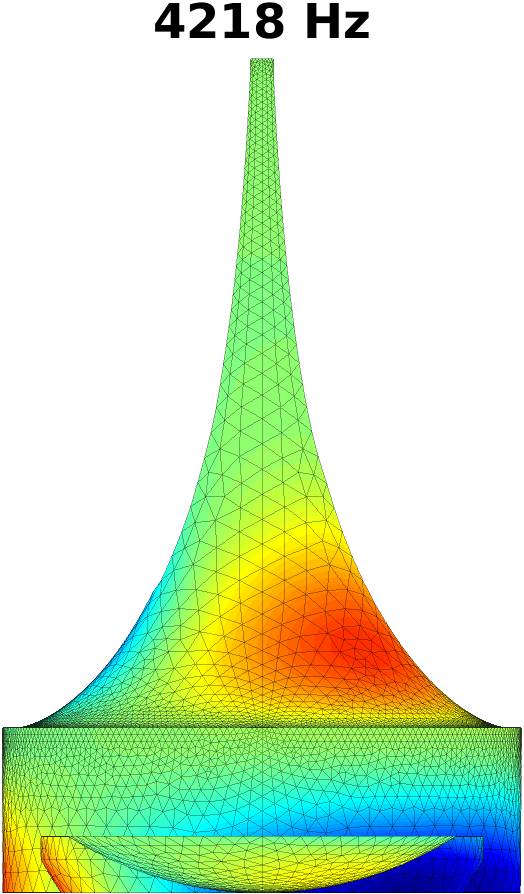} \hspace{0.2cm}
\includegraphics[width=0.22\textwidth]{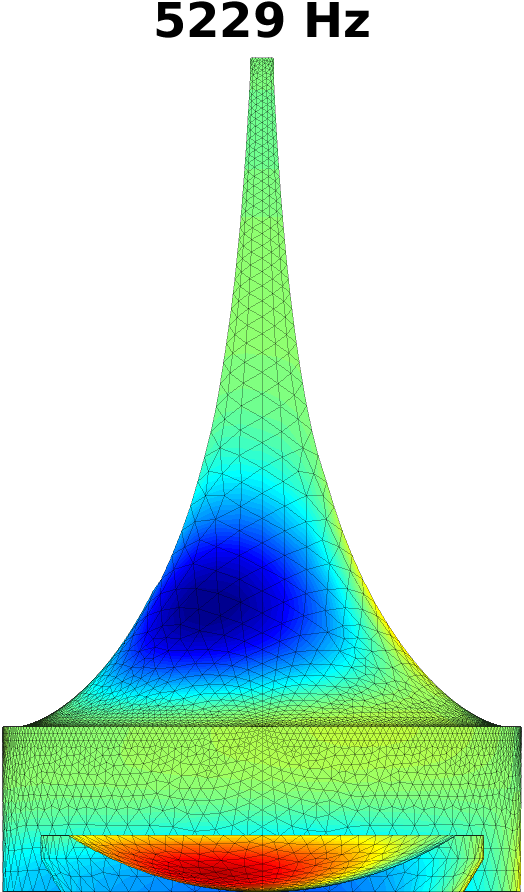}
\end{center}
  \caption{\label{HornSystemRes} Pressure distributions of some
    resonance modes of the impedance matching cavity above the loudspeaker
    unit. The lowest mode is at $1648 \, \textrm{Hz}$, and it is
    purely longitudinal. The lowest transversal mode is at $1994 \,
    \textrm{Hz}$, and it is due to the cylindrical part joining the
    loudspeaker unit to the tractrix horn.  At $4218 \, \textrm{Hz}$,
    a transversal mode appears where strong excitation exists between
    the cylindrical part and the horn.  The lowest transversal mode
    that is solely due to the tractrix horn geometry is found at $5229
    \, \textrm{Hz}$.}
\end{figure}

The purely longitudinal acoustic modes were found at frequencies
$1648 \, \mathrm{Hz}$, $2540 \, \mathrm{Hz}$, $3350 \, \mathrm{Hz}$,
$3771 \, \mathrm{Hz}$, $4499 \, \mathrm{Hz}$, $5061 \, \mathrm{Hz}$,
$5745 \, \mathrm{Hz}$, $6671 \, \mathrm{Hz}$, $7088 \, \mathrm{Hz}$,
$7246 \, \mathrm{Hz}$, and $7737 \, \mathrm{Hz}$.  All of these
longitudinal modes have multiplicity $1$.  Transversal modes divide
into three classes: \textrm{(i)} those where excitation is mainly in
the cylindrical part of the impedance matching cavity, \textrm{(ii)}
those where the excitation is mainly in the tractrix horn, and
\textrm{(iii)} those where both parts of the impedance matching cavity
are excited to equal extent.  Resonances due to the cylindrical part
appear at frequencies $1994 \, \mathrm{Hz}$, $3094 \, \mathrm{Hz}$,
$4150 \, \mathrm{Hz}$, $6063 \, \mathrm{Hz}$, $6262 \, \mathrm{Hz}$,
$6872 \, \mathrm{Hz}$, $6942 \, \mathrm{Hz}$, $7334 \, \mathrm{Hz}$,
and $7865 \, \mathrm{Hz}$, and they all have multiplicity $2$ except
the resonance at $4150 \, \mathrm{Hz}$ that is simple.  (Note that
there is a longitudinal resonance at $4150 \, \mathrm{Hz}$ as well.)
There are only four frequencies corresponding to the transversal modes
(all with multiplicity $2$) in the tractrix horn: namely,
$5229 \, \mathrm{Hz}$, $5697 \, \mathrm{Hz}$, $6764 \, \mathrm{Hz}$,
and $6781 \, \mathrm{Hz}$. The peculiar mixed modes of the third kind
were observed at $4218 \, \mathrm{Hz} \, (2)$,
%$5193 \, \mathrm{Hz} \, (1)$, 
and $5200 \, \mathrm{Hz} \, (3)$ where
the number in the parenthesis denotes the multiplicity.

Based on these observations, the acoustic design of the impedance
matching cavity was deemed satisfactory as the transversal dynamics of
the tractrix horn shows up only above $5.2 \, \mathrm{kHz}$.  The
lower resonant frequencies of the wide end of the cavity are treated
by placement of attenuating material as described in
Section~\ref{ConstructionDetailsSec}.

\subsection{\label{ConstructionDetailsSec} Details of the construction}

%The physical construction of the sound source was carried out taking
%following the lines of reasoning presented above.
%into consideration of the resonant tendencies discovered from the
%Helmholtz model explained above.

The tractrix horn geometry was produced using the parametric Tractrix
Horn Generator OpenSCAD script \cite{TractrixGenerator:2014}. The horn
was 3D printed by Ultimaker Original in PLA plastic with wall thickness
of $2 \, \mathrm{mm}$ and fill density of $100 $\%. The inside surface
of the print was coated by several layers of polyurethane lacquer,
after which it was polished. The horn was installed inside a cardboard
tube, and the space between the horn and the tube was filled with
$\approx 1.2 \, \textrm{kg}$ of \emph{plaster of Paris} in order to
suppress the resonant behaviour of the horn shell itself and to
attenuate acoustic leakage through the horn walls.

The walls of the cylindrical part of the impedance matching cavity
were covered by felt in order to control the standing waves in the
cylindrical part of the cavity. Acoustically soft material, i.e.,
polyester fibre, was placed inside the source (partly including the
volume of the tractrix horn) by the method trial and improvement,
based on iterated frequency response measurements as explained in
Section~\ref{CompensationSubSec} and heuristic reasoning based on
Fig.~\ref{HornSystemRes}.  The main purpose of this work was to
suppress overly strong transversal modes shown in
Fig.~\ref{HornSystemRes} in the impedance matching cavity shown in
Fig.~\ref{DesignOfSourceFig} (middle panel). As a secondary
effect, also the purely longitudinal modes got suppressed. Adding
sound soft material resulted in the attenuation of unwanted resonances
at the cost of high but tolerable increase in the TL of the source.

The loudspeaker unit of the source is contained in the hardwood box
shown in Fig.~\ref{TractrixHornFig}, and its wall thickness $40 \,
\textrm{mm}$. The box is sealed air tight by applying silicone mass to
all joints from inside in order to reduce acoustic leakage. Its exterior
dimensions are $215 \, \textrm{mm} \times 215 \, \textrm{mm} \times
145 \, \textrm{mm}$, and it fits tightly to the horn assembly
described above. The horn assembly and the space of the loudspeaker
unit above the loudspeaker cone form the impedance matching cavity of
the source shown in Fig.~\ref{DesignOfSourceFig}.  There is another
acoustic cavity under the loudspeaker unit whose dimension are $135.0
\, \mathrm{mm} \times 135.0 \mathrm{mm} \times 70.0 \mathrm{mm}$. Also
this cavity was tightly filled with acoustically soft material to
reduce resonances.

\subsection{Electronics and software for measurements}

We use a $4''$ two-way loudspeaker unit (of generic brand) whose
diameter determines the opening of the tractrix horn. Its nominal
maximum output power is $30 \, \mathrm{W \, (RMS)}$ when coupled to a
$4 \, \Omega$ source. The loudspeaker is driven by a power amplifier
based on TBA810S IC. There is a decouplable mA-meter in the
loudspeaker circuit that is used for setting the output level of the
amplifier to a fixed reference value at $1 \, \mathrm{kHz}$ before
measurements. The power amplifier is fed by one of the output channels
of the sound interface ``Babyface'' by RME, connected to a laptop
computer via USB interface.

The acoustic source contains an electret \emph{reference microphone}
(of generic brand, $\oslash \, 9 \, \mathrm{mm}$, biased at $5 \,
\mathrm{V}$) at the output end of the horn. The reference microphone
is embedded in the wave\-guide wall, and there is an aperture of
$\oslash \, 1 \, \mathrm{mm}$ in the wall through which the microphone
detects the sound pressure. The narrow aperture is required so as not
to overdrive the microphone by the very high level of sound at the
output end of the horn, and it is positioned about $13.5 \,
\mathrm{mm}$ below the position where vocal folds would be in the 3D
printed VT model (depending on the anatomy).

The measurements near the mouth position of 3D-printed VTs are carried
out by a \emph{signal microphone}. As a signal microphone, we use
either a similar electret microphone unit as the reference microphone
or Br\"uel \& Kj\ae{}ll measurement microphone model 4191 with the
capsule model 2669 (as shown in Fig.~\ref{TractrixHornFig} (left
panel)) and preamplifier Nexus 2691.  The B\&K unit has over $20 \,
\mathrm{dB}$ lower noise floor compared to electret units which,
however, has no significance when measuring, e.g., the resonant
frequencies of an acoustic load in a noisy environment.  Measurements
in the anechoic chamber yield much cleaner data when the B\&K unit is
used, and this is advisable when studying acoustic loads with higher
TL and lower signal levels. Then, extra attention has to be paid to
all other aspects of the experiments so as to achieve the full
potential of the high-quality signal microphone.
%% (such as quite constricted vowel [\textipa{i}]
%% geometries compared to much wider geometries of
%% [\textipa{\textscripta}].

The reference and the signal electret microphone units were picked
from a set of $10$ units to ensure that their frequency responses
within $80 \, \mathrm{Hz}  \ldots 8 \ \textrm{kHz}$ are practically
identical. It was observed that there are very little differences in
the frequency and phase response of any two such microphone
units. Furthermore, these microphones are practically
indistinguishable from the Panasonic WM-62 units (with nominal
sensitivity $-45 \pm 4$ dB re $1$ V/Pa at $1$ kHz) that were used in
the instrumentation for MRI/speech data acquisition reported in
\cite{A-A-H-J-K-K-L-M-M-P-S-V:LSDASMRIS}.

Final results given in Section~\ref{MeasSigSec} were measured using
the Br\"uel \& Kj\ae{}ll model 4191 at the mouth position. The results
shown in \cite[Fig.~5]{K-M-O:PPSRDMRI} were measured using the
electret unit matched with the similar reference microphone, embedded
to the source at the glottal position. In this article, the electret
microphone measurements at the mouth position were only used for
comparison purposes.

Biases for both the electret microphones are produced by a custom
preamplifier having two identical channels based on LM741 operational
amplifiers. The amplifier has nonadjustable $40 \, \mathrm{dB}$
voltage gain in its passband that is restricted to $40 \, \textrm{Hz}
\ldots 12 \, \textrm{kHz}$. Particular attention is paid to reducing
the ripple in the microphone bias as well as the cross-talk between
the channels. The input impedance $2.2 \, \mathrm{k \Omega}$ of the
preamplifier is a typical value of electret microphones,
% (of which there is no manufacturer's data available),
and the output is matched to $300 \, \Omega$ for the two input
channels of the Babyface unit.

Signal waveforms and sweeps are produced numerically as explained in
Sections~\ref{CalibrationSec} for all experiments.  Frequency response
equalisation and other kinds of time and frequency domain
precompensations are a part of this process. All computations are done
in MATLAB (R2016b) running on Lenovo Thinkpad T440s, equipped with 3.3
GHz Intel Core i7-4600U processor and Linux operating system. The
experiments are run using MATLAB scripts, and access from MATLAB to
the Babyface is arranged through Playrec (a MATLAB
utility,~\cite{Playrec}).

\subsection{Measurement arrangement}

\begin{figure}[t]
\begin{center}
\includegraphics[width=\textwidth]{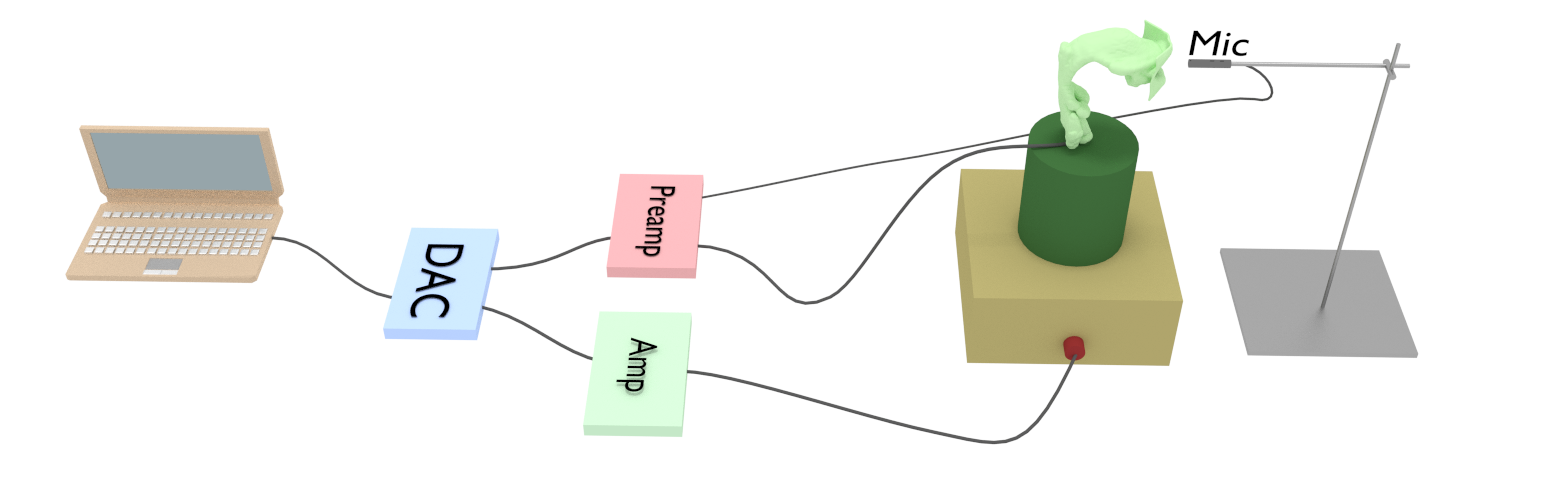}\hspace{0.2cm}\end{center}
  \caption{\label{SystemGraph} An illustration of the
    measurement system.}
\end{figure}

An outline of the measurement arrangement for sweeping a VT print is
shown in Fig.~\ref{SystemGraph}.  Both the amplifiers, the digital
analogue converter (DAC), and the computer are located outside the
anechoic chamber.  The arrangement inside the anechoic chamber
contains two microphones: the reference unit at the glottal position
inside the source, and the external microphone in front of the mouth
opening. The position of the external microphone must be kept same in
all measurements to have reproducibility.

Because of the quite high transmission loss of the VT print (in
particular, in VT configuration corresponding to [\textipa{i}]) and
the relatively low sound pressure level produced by the source at the
glottal position (compared to the sound pressure produced by human
vocal folds), one may have to carry out measurements using an acoustic
signal level only about $20 \ldots 30 \, \mathrm{dB}$ above the
hearing threshold. The laboratory facilities require using
well-shielded coaxial microphone cables of length $10 \, \mathrm{m}$
in order to prevent excessive hum. Another significant source of
disturbance is the acoustic leakage from the source directly to the
external microphone. This leakage was be reduced by $\approx 6 \,
\mathrm{dB}$ by enclosing the sound source into a box made of
insulating material, and preventing sound conduction through
structures by placing the source on silicone cushions resting on a
heavy stone block (not shown in
Figs.~\ref{TractrixHornFig}~and~\ref{SystemGraph}).

\section{\label{CompValSec} Computational validation using a VT load}

When an acoustic load is coupled to a sound source containing an
impedance matching cavity, the measurements carried out using the
source necessarily concern the joint acoustics of the source and the
load. Hence, precautions must be taken to ensure that the
characteristics of the acoustic load truly are the main component in
measurement results. In the case of the proposed design, the small
intersectional area of the opening at the source output leads to high
acoustic output impedance which is consistent with a reasonably good
acoustic \emph{current} source. Also, the narrow glottal position of
the VT helps in isolating the the two acoustic spaces from each other.

\begin{figure}[]
\begin{center}
\includegraphics[width=0.22\textwidth]{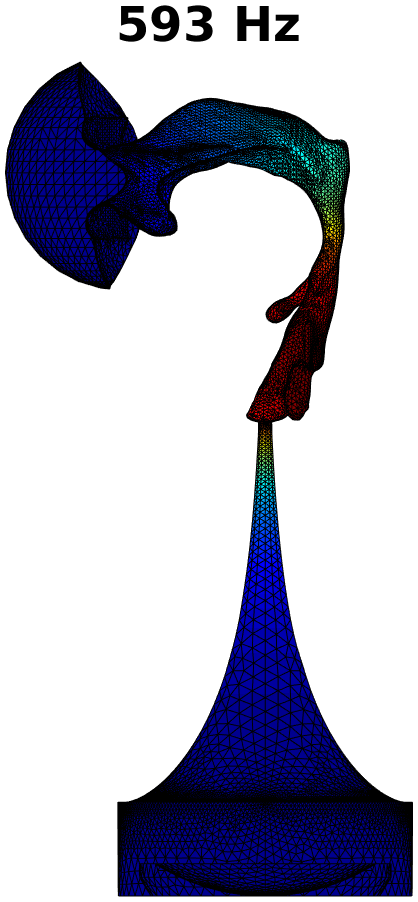}\hspace{0.2cm}
\includegraphics[width=0.22\textwidth]{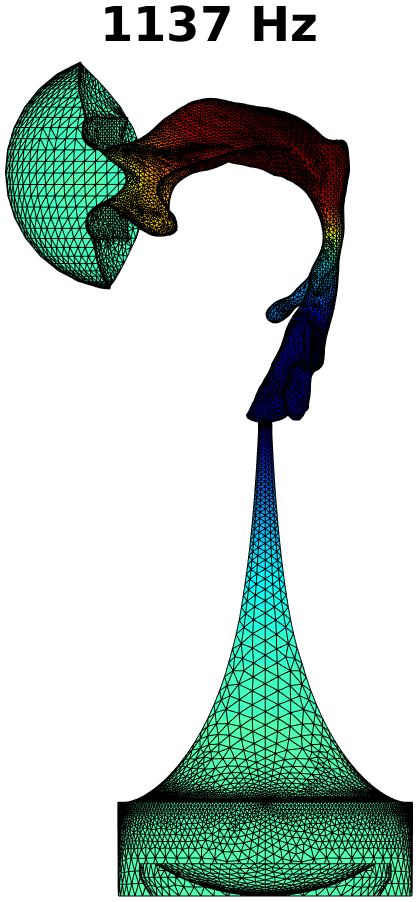} \hspace{0.2cm}
\includegraphics[width=0.22\textwidth]{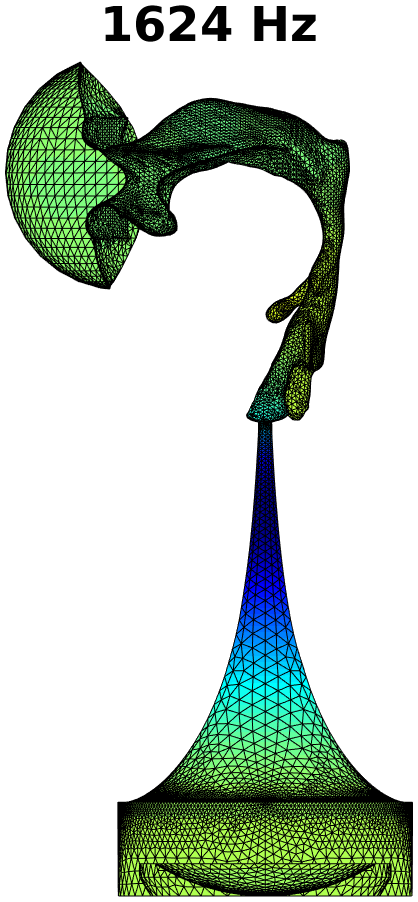}  \vspace{0.5cm}
\includegraphics[width=0.22\textwidth]{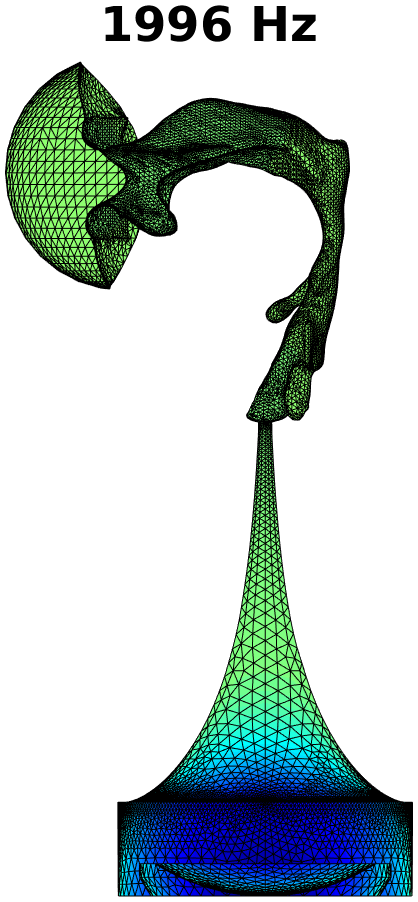} \\ \hspace{0.2cm}
\includegraphics[width=0.22\textwidth]{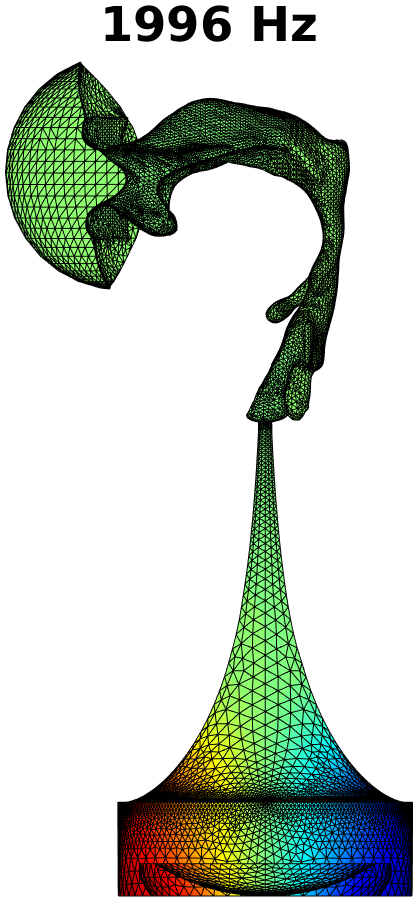}\hspace{0.2cm}
\includegraphics[width=0.22\textwidth]{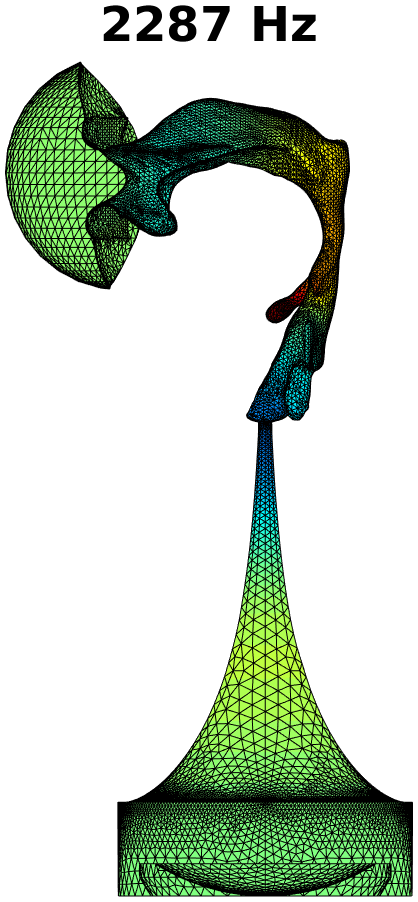} \hspace{0.2cm}
\includegraphics[width=0.22\textwidth]{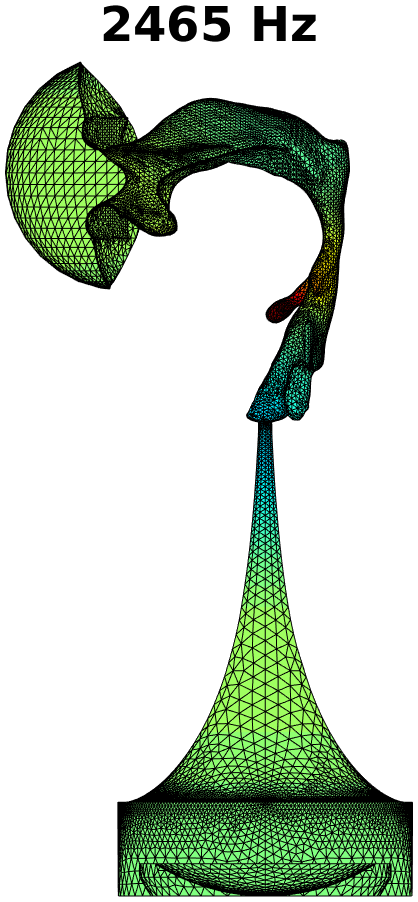} \hspace{0.2cm}
\includegraphics[width=0.22\textwidth]{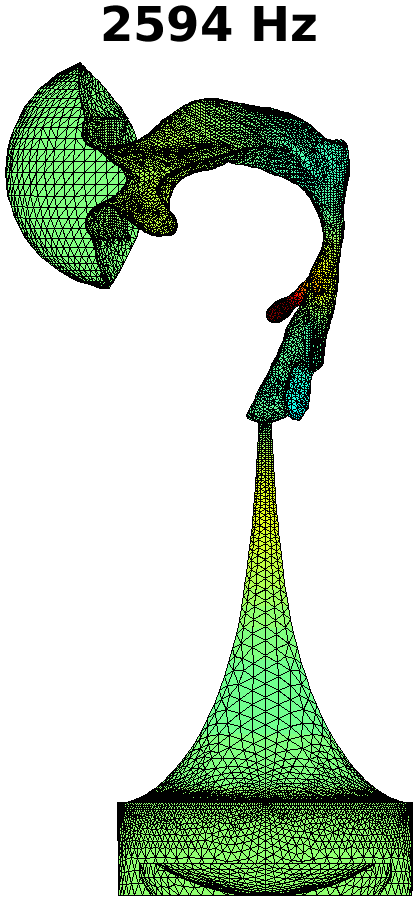}
\end{center}
  \caption{\label{CoupledSystemRes} Pressure distributions of some
    resonance modes of the impedance matching cavity of the source
    coupled to a VT geometry of [\textipa{\textscripta}]. The modes
    corresponding to longitudinal VT resonances are at
    frequencies $593 \, \textrm{Hz}$, $1137 \, \textrm{Hz}$, and $2287
    \, \textrm{Hz}$, corresponding to formants $F_1, F_2,$ and $F_3$.
    The remaining pressure modes under $2465 \, \mathrm{Hz}$ are
    excitations of the impedance matching cavity of the source.}
\end{figure}

We proceed to evaluate this isolation by computing the Helmholtz
resonance structures of the joint system shown in
Figs.~\ref{CoupledSystemRes} and compare them with \textrm{(i)}
formant frequencies measured from the same test subject during the MR
imaging, and \textrm{(ii)} Helmholtz resonances of the VT geometry
shown in Fig.~\ref{DesignOfSourceFig} (right panel).  The VT part of
both the computational geometries is the same, and it corresponds to
the vowel [\textipa{\textscripta}]. The vowel [\textipa{\textscripta}]
out of [\textipa{\textscripta, i, u}] was chosen because its three
lowest formants are most evenly distributed in the voice band of
natural speech.

\begin{table}[h]
 \centerline{
    \begin{tabular}{|l|c|c|c|c|}
\hline
 & $F_1$  & $F_2$  & $F_3$  \\
\hline
\textbf{VT resonances} & $519$ & $1130$ & $2297$ \\
\textbf{VT + source resonances} & $594$ & $1136$ &  $2290$ \\
\textbf{Formant frequencies}  & $683$ & $1111$ &  $2417$ \\
\hline
\end{tabular}}
\caption{\label{FormantCouplingTable} Vowel formants and Helmholtz
  resonances (in $\textrm{Hz}$) of a VT during a production
  of [\textipa{\textscripta}]. In the first two row, only those
  resonances have been taken into account whose modal behaviour
  corresponds with the formants $F_1, F_2,$ and
  $F_3$.} % \DisplayNote{JM: how were the formants obtained?}}
\end{table}

In numerical computations, the domain $\Omega \subset \R^3$ for the
Helmholtz equation~\eqref{HelmHoltzEq} consists of the VT geometry of
[\textipa{\textscripta}] either as such (leading to ``VT resonances''
in Table~\ref{FormantCouplingTable}) or manually joined to the
impedance matching cavity at the glottal position (leading to ``VT +
source resonances'' in Table~\ref{FormantCouplingTable}).  The FEM
meshes have been described in Section~\ref{HelmholtzCavitySec}.
%% The  details of the FEM meshes are given in Table~\ref{MeshTable}. 
The acoustic modes and resonant frequencies have been computed from
Eq.~\eqref{HelmHoltzEq}, and some of the resulting resonant
frequencies and modal pressure distributions are shown in
Fig.~\ref{CoupledSystemRes}.

In contrast to Section~\ref{HelmholtzCavitySec}, the symbol $\Gamma_0$
now denotes the spherical mouth interface surface visible in
Fig.~\ref{DesignOfSourceFig} (right panel), and instead of
Eq.~\eqref{DirichletBndry} we use the boundary condition of Robin type
\begin{equation*}
    \lambda \phi_\lambda({\bf r}) + c\frac{\partial
      \phi_\lambda}{\partial\nu}({\bf r}) = 0 \text{ on } \Gamma_0,
\end{equation*}
making the interface absorbing. When computing VT resonances for
comparison values without the impedance matching cavity (the top row
in Table~\ref{FormantCouplingTable}), the interface at the glottal
opening is considered as part of $\Gamma_0$, too.  The resulting
quadratic eigenvalue problem was then solved by transforming it to a
larger, linear eigenvalue problem as explained in
\cite[Section~3]{Hannukainen:2007}.
% The computed resonant frequencies are given in Table~\ref{FormantCouplingTable}. 
For a similar kind of numerical experiment involving VT geometries but
without a source, see \cite{arnela:2013}.

The formant values given in Table~\ref{FormantCouplingTable} have been
extracted by Praat \cite{Praat:2016} from post-processed speech
recordings during the acquisition of the MRI geometry as explained in
Section~\ref{ProcessingSubSec}. The extraction was carried out at $3.5
\, \mathrm{s}$ from starting of the phonation, with duration $25 \,
\textrm{ms}$.

Given in semitones, the discrepancies between the first two rows in
Table~\ref{FormantCouplingTable} are $-2.3$, $-0.1$, and
$0.05$. Similarly, the discrepancies between the last two rows in
Table~\ref{FormantCouplingTable} are $-2.4$, $0.4$, and $-0.9$.  The
largest discrepancy concerning the first formant $F_1$ is partly
explained by the challenges in formant extraction from the nonoptimal
speech sample pair of the MRI data used. In
\cite[Table~2]{A-A-H-J-K-K-L-M-M-P-S-V:LSDASMRIS}, the value for $F_1$
from the same test subject was found to be $580 \pm 23 \, \mathrm{Hz}$
based on averaging over ten speech samples during MRI and using a more
careful treatment for computing the spectral envelope, based on MATLAB
function \verb|arburg| .

We conclude that for Helmholtz resonances corresponding to $F_2$ and
$F_3$ of the physical model of [\textipa{\textscripta}], the
perturbation due to acoustic coupling with the impedance matching
cavity are small fractions of the comparable natural variation in
spoken vowels. So as to the lowest formant $F_1$, it seems that the
impedance matching cavity actually represents a better approximation
of the true subglottal acoustics contribution than the mere absorbing
boundary condition imposed at the glottis position of a VT
geometry. We further observe that the three lowest resonant modes of
the VT (corresponding to formants $F_1, F_2, F_3$) appear where the
impedance matching cavity remains in ``ground state''; see
Fig.~\ref{CoupledSystemRes}. This supports the desirable property that
the narrowing of the horn at the vocal folds position effectively
keeps the impedance matching cavity of the source and the VT load only
weakly coupled.

% \newpage

\section{Calibration measurements and source compensation}
\label{CalibrationSec}

\subsection{\label{CompensationSubSec} Measurement and compensation of the frequency response}

\begin{figure}[t]
  \includegraphics[width=0.32\textwidth]{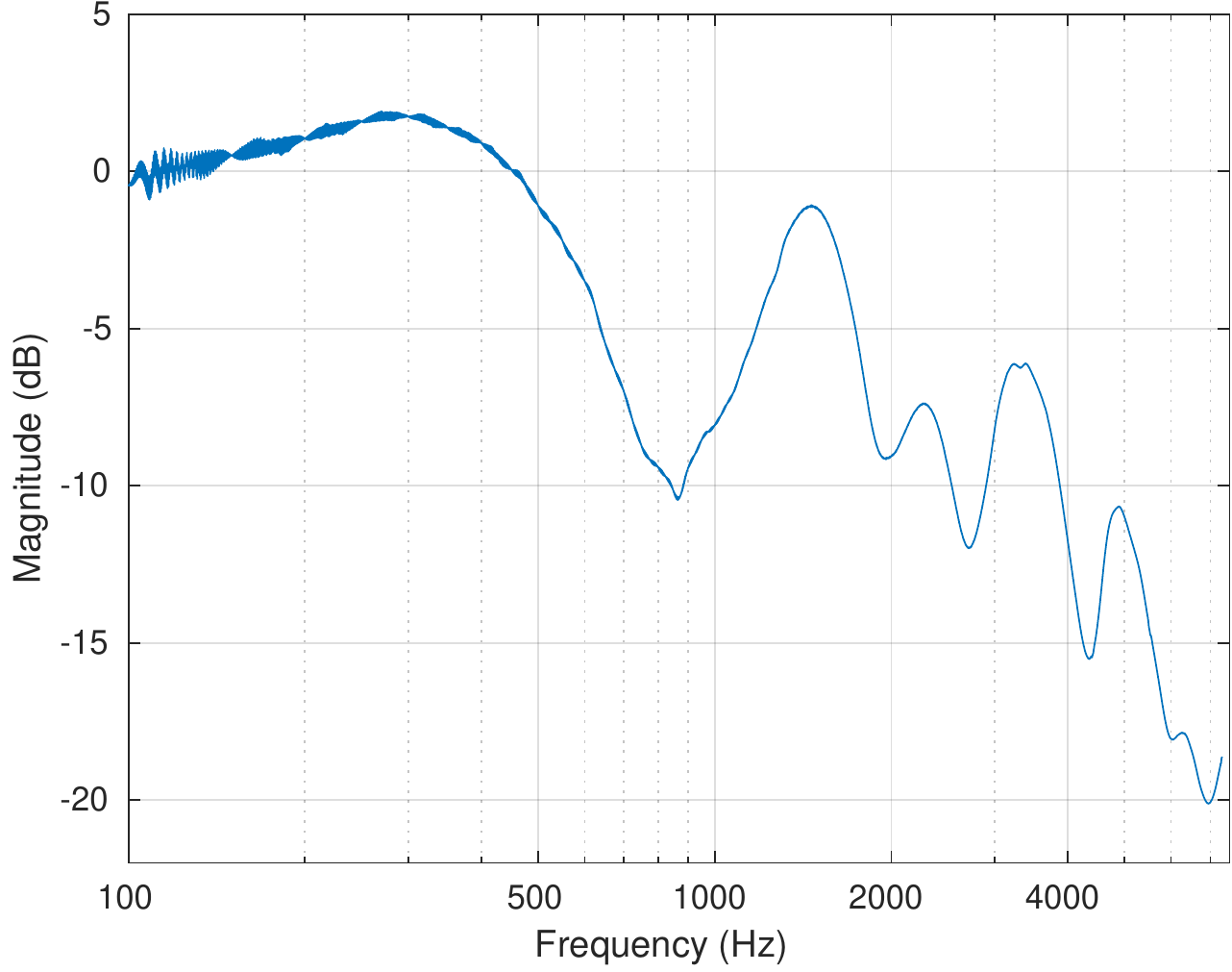}
  \includegraphics[width=0.32\textwidth]{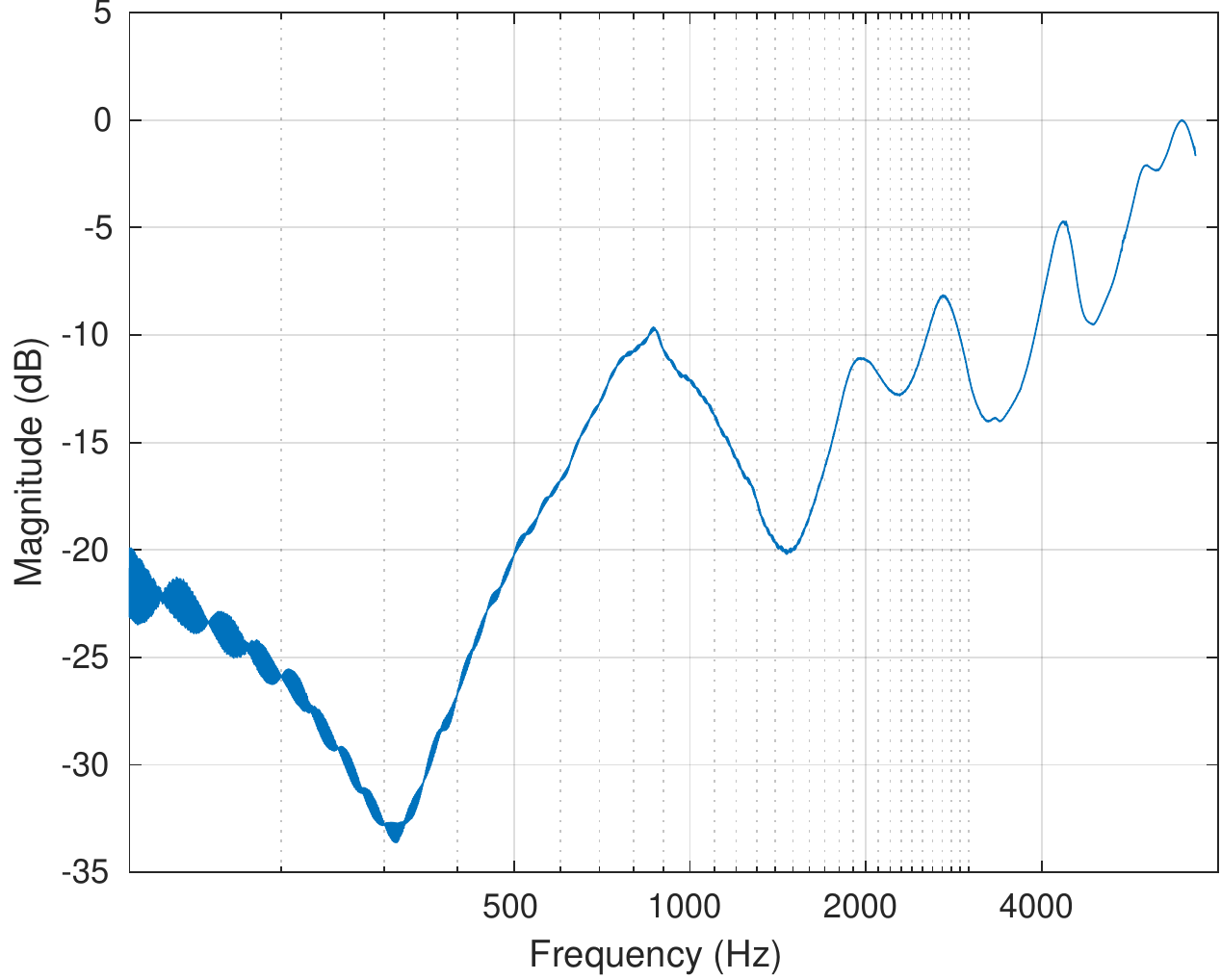}
  \includegraphics[width=0.32\textwidth]{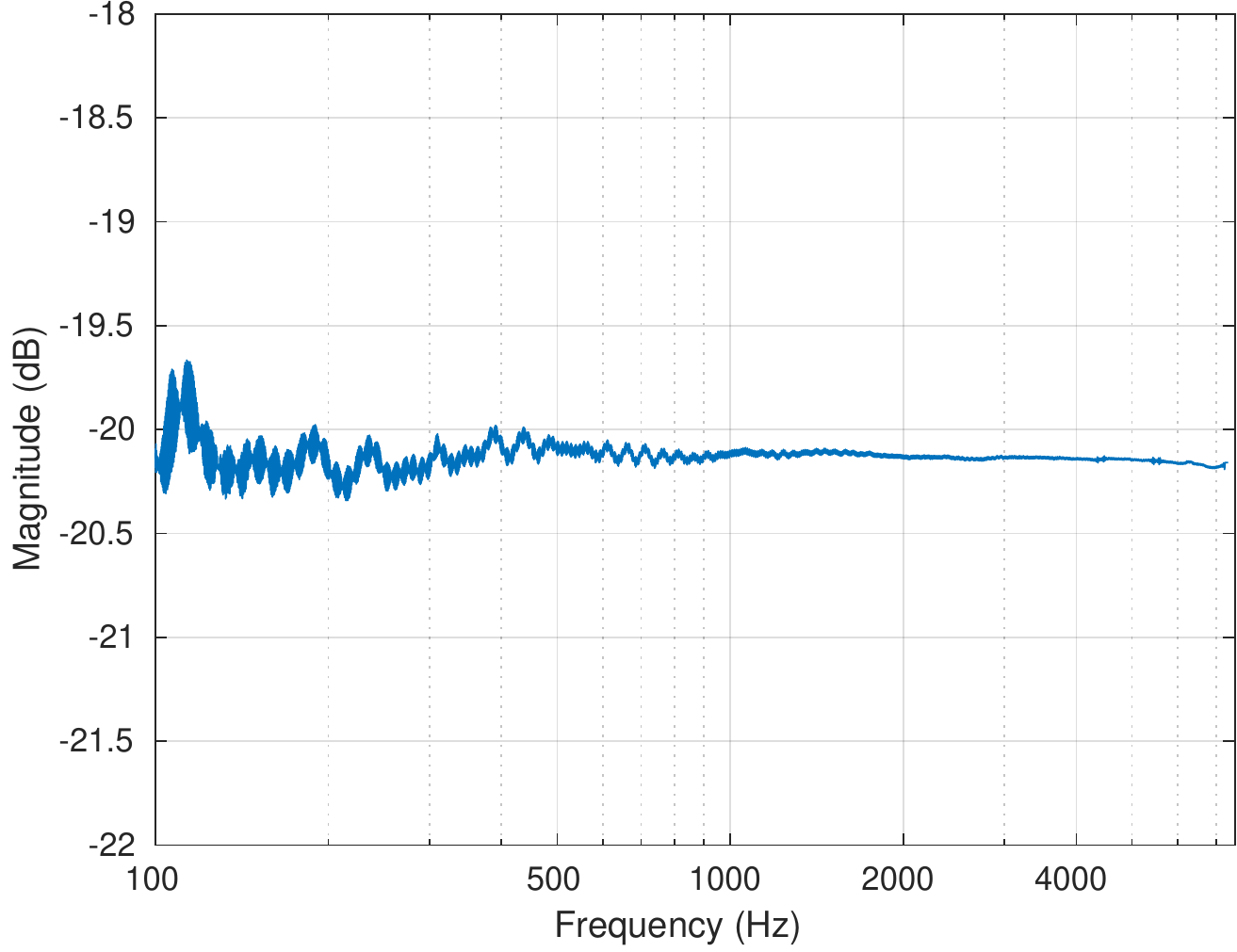}
  \caption{\label{EnvelopeResidualFig} Left panel: The pressure signal
    envelope of the measurement system at the glottal position when a
    constant amplitude exponential voltage sweep was used as the
    loudspeaker input. The first longitudinal resonance of the
    impedance matching cavity appears at $1648 \, \mathrm{Hz}$. The
    source was terminated to the dummy load shown in
    Fig.~\ref{TractrixHornFig} (right panel).  Middle panel: The
    inverse weights that are applied to the constant amplitude
    exponential sweep in order to get the output in the next panel.
    Right panel: The envelope of the weighted exponential sweep at the
    glottal position where the weight has been produced by
    Algorithm~\ref{algorithm}.  The produced sound pressure sweep at
    the source output has residual amplitude dynamics of approximately
    $0.5 \,\mathrm{dB}$.}
\end{figure}

\begin{figure}[hb]
  \centering
  \includegraphics[width=0.3\textwidth]{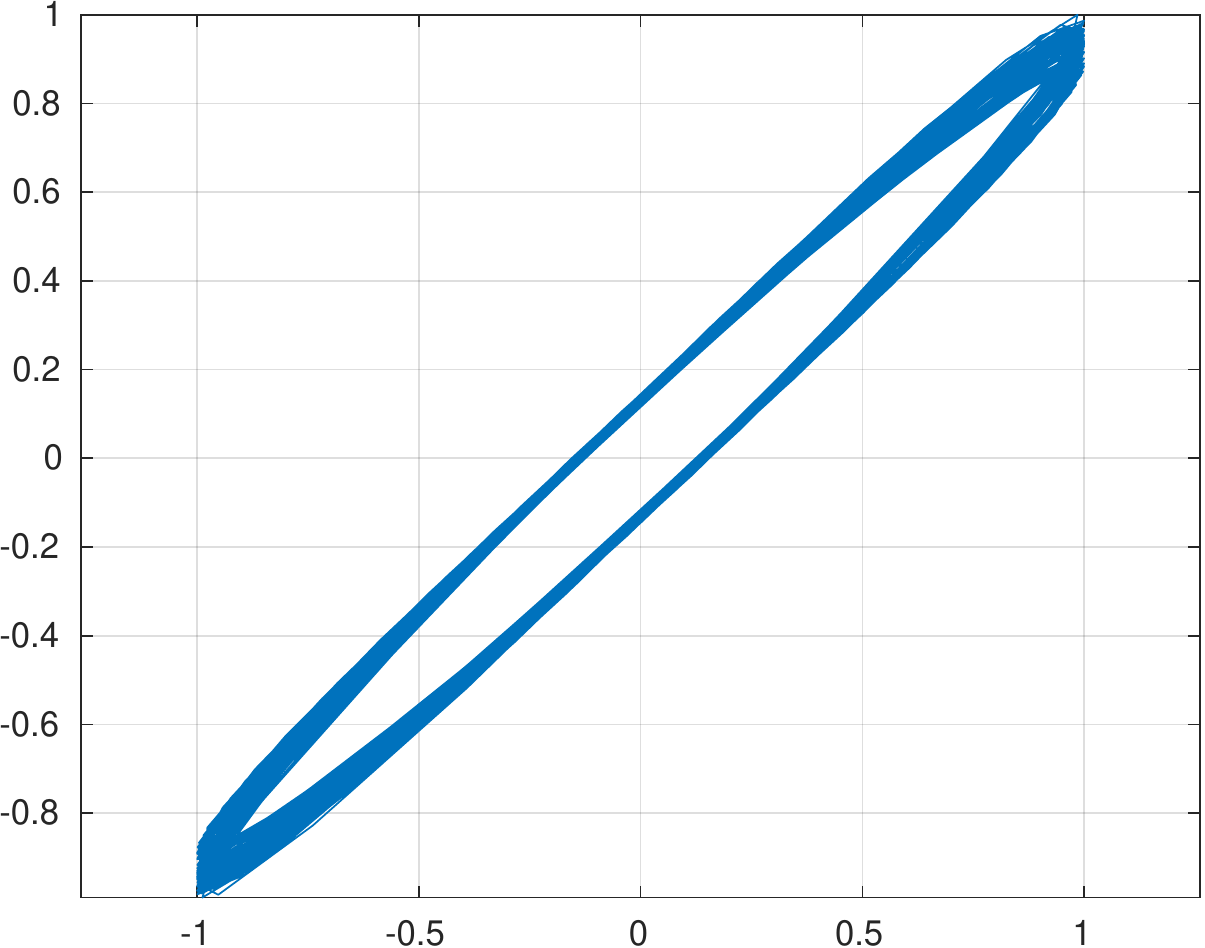} \hspace{0.2cm}
  \includegraphics[width=0.3\textwidth]{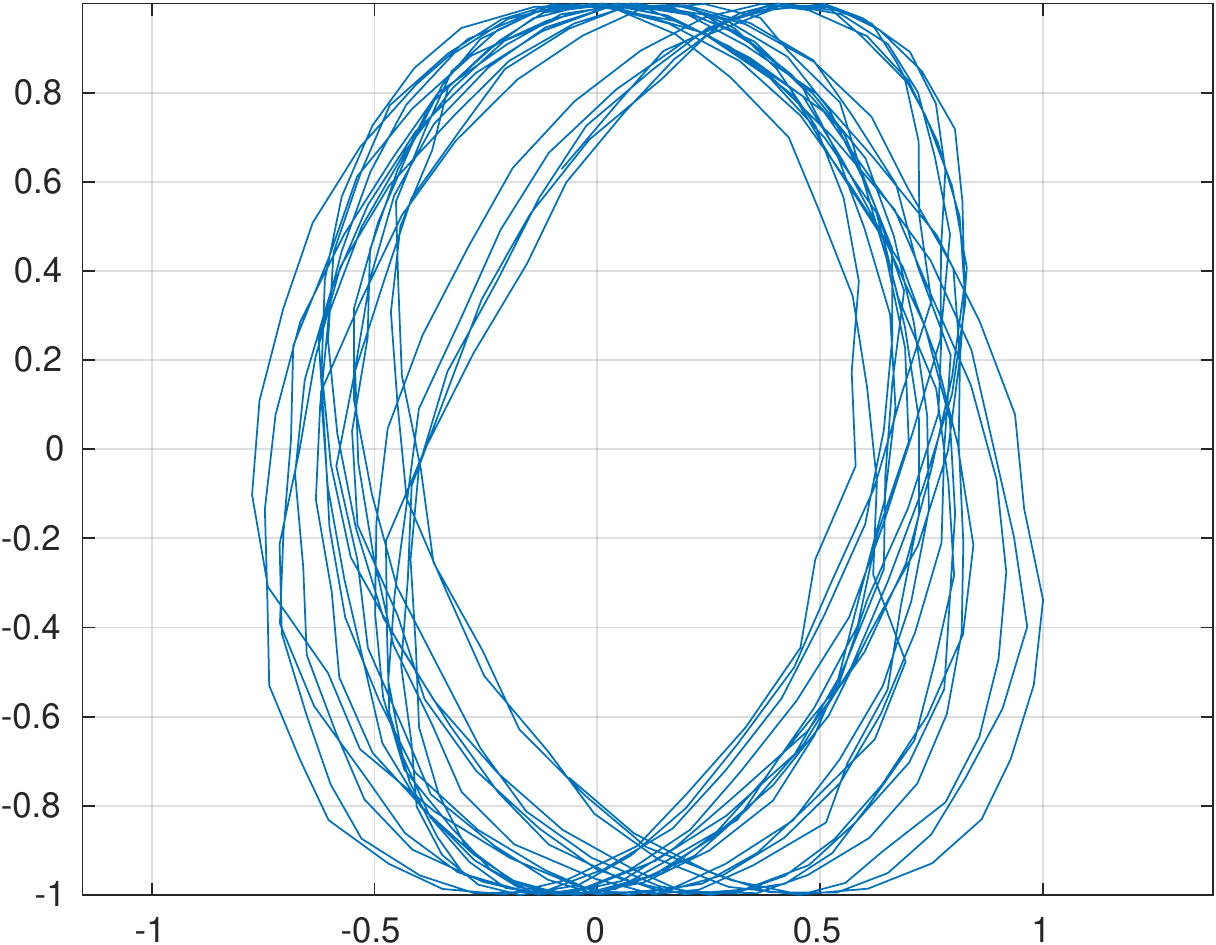} \hspace{0.2cm} 
  \includegraphics[width=0.33\textwidth]{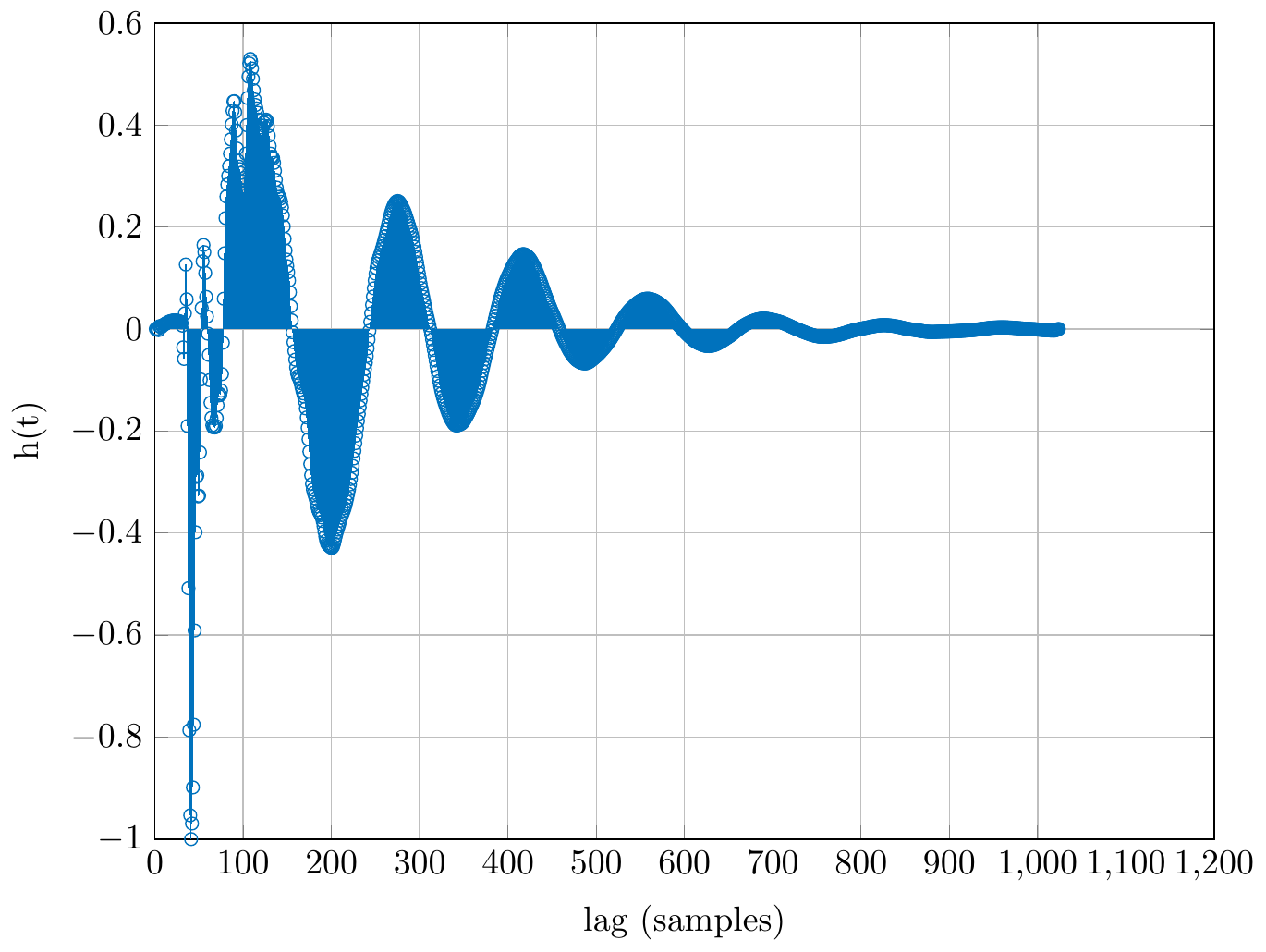}
\caption{\label{LissajousFig} Left: Lissajous plot of the original,
  unweighted voltage sweep against the sweep near $3 \, \mathrm{kHz}$
  weighted by $\mathbf{w}$ produced by
  Algorithm~\ref{SweepAlgorithm}. Middle: Lissajous plot of the
  unweighted voltage sweep against the corresponding output as
  recorded by the reference microphone near $1540 \mathrm{Hz}$ where
  the phase difference varies around $\pi/2$. Right: The measured
  impulse response of from the voltage input to reference microphone
  output. In both the measurements, the source was terminated to the
  dummy load shown in Fig.~\ref{TractrixHornFig} (right panel).}
\end{figure}

In this section, we describe the production of an \emph{exponential
  frequency sweep}\footnote{Also known as the logarithmic chirp.} with
uniform sound pressure at the glottal position. The defining property
of such sweeps is that each increase in frequency by a semitone takes
an equal amount of time. In this work, the frequency interval of such
sweeps is $80 \ldots 7350 \, \mathrm{Hz}$ with duration of $10 \,
\mathrm{s}$. All measurements leading to curves in
Figs.~\ref{EnvelopeResidualFig}--\ref{LissajousFig} were carried out
using the \emph{dummy load} shown in Fig.~\ref{TractrixHornFig} (right
panel) as the standardised acoustic reference load.

If one plainly introduces a constant voltage amplitude exponential
sinusoidal sweep to the loudspeaker unit, the sound pressure at the
source output (as seen by the adjacent reference microphone) will vary
over $20 \, \mathrm{dB}$ over the frequency range of the sweep as
shown in Fig.~\ref{EnvelopeResidualFig} (left panel). The key
advantage in producing a \emph{constant amplitude sound pressure} at
the source output is that excessive external noise contamination of
measured signals can be avoided on frequencies where the output power
would be low. Standardising the sound pressure at the output of the
source also makes the source acoustics less visible in the
measurements of the load. This reduces the perturbation effect at
$F_1$ that was computationally observed in Section~\ref{CompValSec}.

An essentially flat sound pressure output shown in
Fig.~\ref{EnvelopeResidualFig} (right panel) can be obtained from the
source by applying the frequency dependent amplitude weight
$\mathbf{w}$ shown in Fig.~\ref{EnvelopeResidualFig} (middle panel) to
the voltage input to the loudspeaker unit. As is to be expected, both
the weighted and unweighted voltage sweeps have almost identical phase
behaviours as can be seen in Fig.~\ref{LissajousFig} (left panel).  In
contrast, the voltage sweep and the resulting sound pressure at the
reference microphone are out of phase in a very complicated frequency
dependent manner; see Fig.~\ref{LissajousFig} (middle panel). Such
phase behaviour cannot be explained by the relatively sparsely located
acoustic resonances of the impedance matching cavity.

An iterative process requiring several sweep measurements was devised
to obtain the weight shown in Fig.~\ref{EnvelopeResidualFig} (middle
panel), and it is outlined below as
Algorithm~\ref{SweepAlgorithm}. Various parameters in the algorithm
were tuned by trial and error so as to produce convergence to a
satisfactory compensation weight. During the iteration, different
versions of the measured sweeps have to be temporally aligned with
each other. The required synchronisation is carried out by detecting a
$1 \, \mathrm{kHz}$ cue of length $1 \, \mathrm{s}$, positioned before
the beginning of each sweep. This is necessary because there are
wildly variable latency times in the DAC/software combination used for
the measurements.

\begin{algorithm}
\caption{Computation of the equalisation weight $\mathbf{w}$}\label{algorithm}
\begin{algorithmic}[1]
\Procedure{CalibrateCompensation}{n,t}
\State $\mathbf{w}\gets [1,1,\ldots,1]$
\For {$k\gets 0\ldots N$}
\State $\mathbf{x} \gets w\cdot $ExponentialChirp(t)
\State $\mathbf{y} \gets $Play $ (\mathbf{x}) $
\State $H \gets $ComputeEnvelope$ (\mathbf{y})$
\State $d \gets $Dynamics $(H)$
\State $r \gets $Regularization $(d)$
\State $ \mathbf{w} \gets \frac{1}{\abs{H}+r}\cdot \mathbf{w}$
\EndFor
\State\Return $\mathbf{w}$
\EndProcedure
\end{algorithmic} \label{SweepAlgorithm}
We consider the calibration successful if the measured dynamics at the
final iteration stage is below $1 \, \mathrm{dB}$.
\end{algorithm}

The system comprising the power amplifier, the loudspeaker and the
acoustic load is somewhat nonlinear which becomes evident in wide
frequency ranges and high amplitude variations. Even though the curves
in Fig.~\ref{EnvelopeResidualFig} (left and middle panels) are
obviously related, they do not sum up to a constant that would be
independent of the frequency.  Not even the dynamical ranges of these
curves coincide as would happen in a linear and time-invariant
setting. In spite of nonlinearity, it is possible to use of a very
slowly increasing sweep to produce an accurate voltage gain from the
output of DAC to the output of reference microphone preamplifier over
a very wide range of frequency. One example of such voltage gain
function is shown in Fig.~\ref{EnvelopeResidualFig} (left panel) but
its inverse is not a good candidate for the compensation weight.

\noindent The results of sweep measurement from physical models of VT
are given in Section~\ref{SweepMeasSubSec}.

\subsection{Compensation of the source response for reference tracking}
\label{ImpulseSubSec}

Another important goal is to be able to reconstruct a desired waveform
as the sound pressure output of the source as observed by the
reference microphone. In the context of speech, a good candidate for a
target waveform is the Liljencrants--Fant (LF) waveform \cite{Fant:1985}
describing the the flow through vibrating vocal folds; see
Fig.~\ref{ReconstructedWaveformFig} (top row, left panel).

Because there is an acoustic transmission delay of $\approx 0.5 \,
\mathrm{ms}$ in the impedance matching cavity in addition to various,
much larger latencies in the DAC/computer instrumentation and software,
a simple feedback-based PID control strategy is not feasible for
solving any trajectory tracking problem. Instead, a \emph{feedforward
  control solution} is required where the response of the acoustic
source and the electronic instrumentation is cancelled out by
\emph{regularised deconvolution}, so as to obtain an input waveform
that produces the desired output. For this, we use the version of
constrained least squares filtering whose mathematical treatment in
signal processing context is given in Section~\ref{DeConvSec}.

The regularised deconvolution requires estimating the impulse response
of the whole measurement system that corresponds to the convolution
kernel $h_0$ in Eq.~\eqref{ConvolutionEq}. This response is estimated
using the sinusoidal sweep excitation described in
\cite{Muller:TFM:2001}, and the result of the measurement can be seen
in~Fig.~\ref{LissajousFig} (right panel). Because the deconvolution
contains regularisation parameters $\gamma$ and $\kappa$, it tolerates
some noise always present in the estimated impulse response.

Let us proceed to describe how the mathematical treatment given in
Section~\ref{DeConvSec} can be turned into a workable signal
processing algorithm in discrete time. All signals (including the
estimated impulse response corresponding to kernel $h_0$) are
discretised at the sampling rate $44 100 \, \mathrm{Hz}$ used in all
signal measurements. We denote the sample number of a discretised
signal, say, $x[n]$ by $N = 44 100 \, \mathrm{Hz} \cdot T$ where $T$
is the temporal length of the original (continuous) signal $x(t)$, $t
\in [0, T]$, and sampling is carried out by setting, e.g., 
\begin{equation*}
  x[n] = \frac{1}{T_s} \int_{(n-1) T_s}^{n T_s}{ x(t) \, dt} 
  \quad \text{ where } \quad 1 \leq n \leq N \quad
  \text{ and } \quad T_s = \mathrm{s}/44 \,  100 .
\end{equation*}
The measured (discrete) impulse response $h_0[n]$ is extended to match
the signal length $N$ by padding it with zeroes, if necessary.

In discrete time, the regularised deconvolution given in
Eqs.~\eqref{NormalEq}--\eqref{ResidualEq} takes the matrix/vector form
\begin{equation} \label{DiscretisedRegConv}
\begin{aligned}
  \v{\mathbf{u}} & = \left(\gamma \left (\kappa I +  R^{T}R \right ) +  H^{T}H  \right)^{-1}H^{T} \mathbf{y} \quad \text{ and } \\
  \mathbf{v}_{\gamma, \kappa} & =
\left(\kappa I +  R^{T}R  + \gamma^{-1} H^{T}H  \right)^{-1}  \left ( \kappa I + R^{T}R  \right ) \mathbf{y}.
\end{aligned}
\end{equation}
The components of the $N \times 1$ column vectors $\v{\mathbf{u}},
\mathbf{y}, \mathbf{v}_{\gamma, \kappa}$ are plainly the discretised
values $\v{u}[n], y[n], {v}_{\kappa, \mu}[n]$ for $n = 1, \ldots , N$
of signals $\v{u}, y, {v}_{\kappa, \mu}$, respectively, given in
Eqs.~\eqref{NormalEq}--\eqref{ResidualEq} where $\mu = \gamma^{-1}$.
The second order difference $N \times N$ matrix $R$ is the symmetric
matrix whose top row is $\left [ 2,-1,0,\ldots,0 , -1 \right ]$,
making it circulant. The nonsymmetric $N \times N$ matrix $H = \left [
  h_{j,k} \right ]$ is constructed by setting $h_{jk}=h_0[(N + j -
  k)\,\mathrm{mod} \,N+1]$ for $1 \leq j, k \leq N$. Because all of
the matrices $R = R^{T}, H, H^{T}$ are now circulant, so is the
symmetric matrix $\gamma \left (\kappa I + R^{T}R \right ) + H^{T}H$
in Eq.~\eqref{DiscretisedRegConv}. Hence, the matrix/vector products
in Eq.~\eqref{DiscretisedRegConv} can be understood as circular
discrete convolutions that can be implemented in $N\log(N)$ time using
the Fast Fourier Transform (FFT). This leads to very efficient
solution for $ \v{\mathbf{u}}$ given $\mathbf{y}$ even for long
signals.

Defining the transfer functions $\widehat R(z)$, $\widehat H(z)$ and
the transforms $\widehat{y}(z), \widehat{v}_{\gamma, \kappa} (z)$ for
$z = e^{i \theta}$ as
\begin{equation*}
\begin{aligned}
  & \widehat R(z) = - z^{-N} - z^{-1} + 2  -  z  - z^{N},  \quad 
  \widehat H(z) = \sum_{n = 0}^N {h_{n 0} z^{n}} + \sum_{n = -N}^{-1} {h_{0 n} z^{n}}, \\
  & \widehat{y}(z) = \sum_{n = 1}^N {y[n] z^{n}}, \text{ and } 
  \widehat{v}_{\gamma, \kappa} (z) = \sum_{n = 1}^N {{v}_{\gamma, \kappa}[n] z^{n}},
\end{aligned}
\end{equation*}
we observe that the latter of Eqs.~\eqref{DiscretisedRegConv} takes
the form of Discrete Fourier Transform (DFT)
\begin{equation} \label{DFTTransferFunctionEq}
  \frac{\widehat{v}_{\gamma, \kappa} (z_k) }{\widehat{y}(z_k)} = \frac{\kappa + \abs{\widehat R(z_k)}^2 }
          {\kappa + \abs{\widehat R(z_k)}^2 + \gamma^{-1} \abs{\widehat H(z_k)}^2 },
\end{equation}
realised in MATLAB code, where $z_k = e^{2 \pi k/N}$ and $k = 1,
\ldots , N$ enumerates the discrete frequencies.  By Parseval's
identity, we interpret the residual equation \eqref{remainderEq} in
discretised form as
\begin{equation*}
\sum_{k = 1}^N {\abs{\widehat{v}_{\gamma, \kappa} (z_k)}^2} = \epsilon^2 \sum_{k = 1}^N {\abs{\widehat{y} (z_k)}^2}
\end{equation*}
which, together with Eq.~\eqref{DFTTransferFunctionEq}, gives an
equation from which $\gamma = \gamma(\epsilon, \kappa)$ can be solved for each $0 < \epsilon < 1$
and $\kappa \geq 0$. This is done using MATLAB's \texttt{fminbnd}
function to ensure that $\gamma>0$.  The values for $\epsilon, \kappa$
are chosen based on the experiments.

\section{\label{MeasSigSec} Results}

Two kinds of measurements on 3D printed VT physical models were
carried out. Firstly, the measurement of the magnitude frequency
response to determine spectral characteristics (such as the lowest
resonant frequencies) of the VT geometry. Secondly, the classical LF
signal was fed into the VT physical model to simulate vowel acoustics
in a spectrally correct manner.

\subsection{\label{SweepMeasSubSec} Sweep measurements}

\begin{figure}[h]
  \centering
  \includegraphics[width=0.3\textwidth]{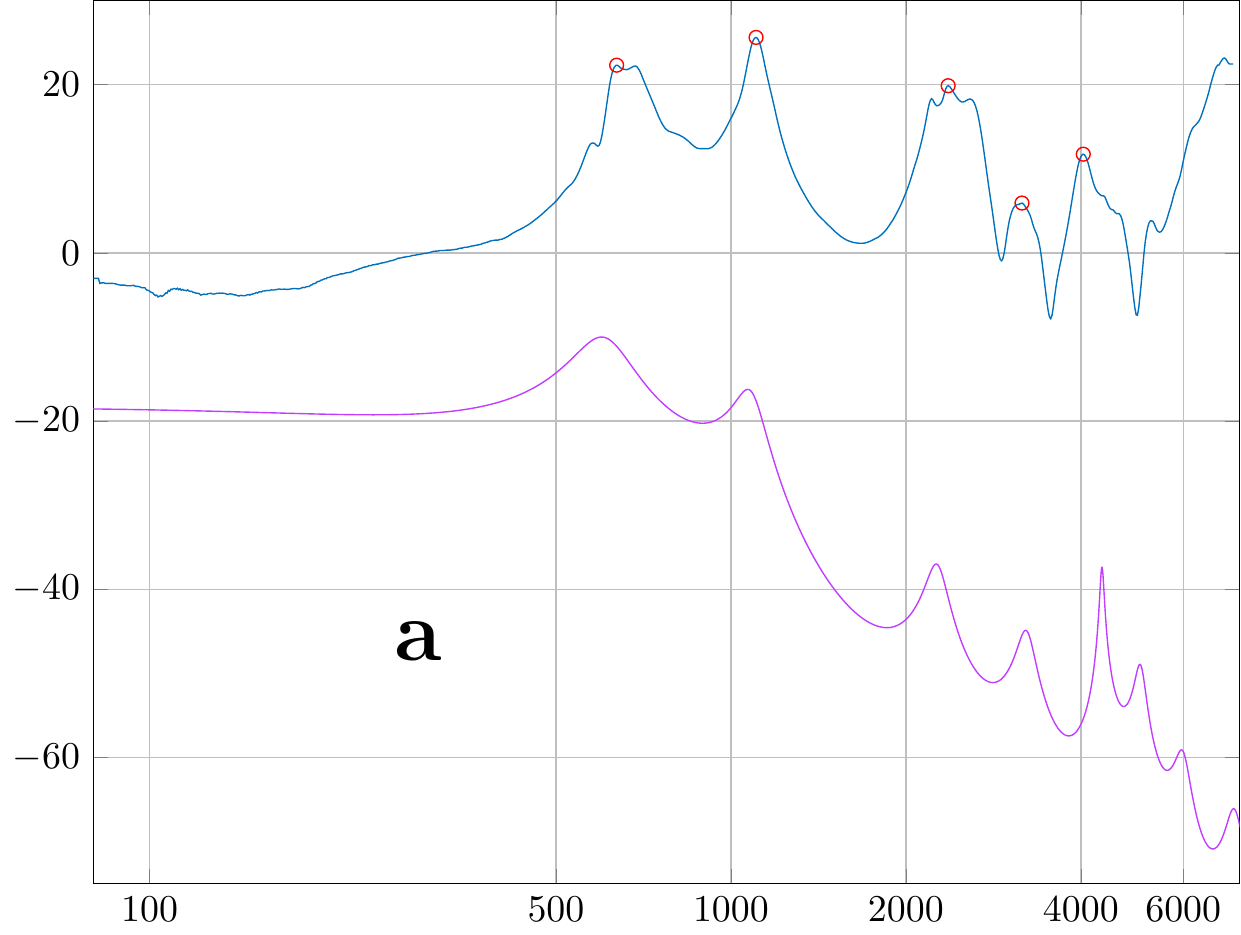}
  \includegraphics[width=0.3\textwidth]{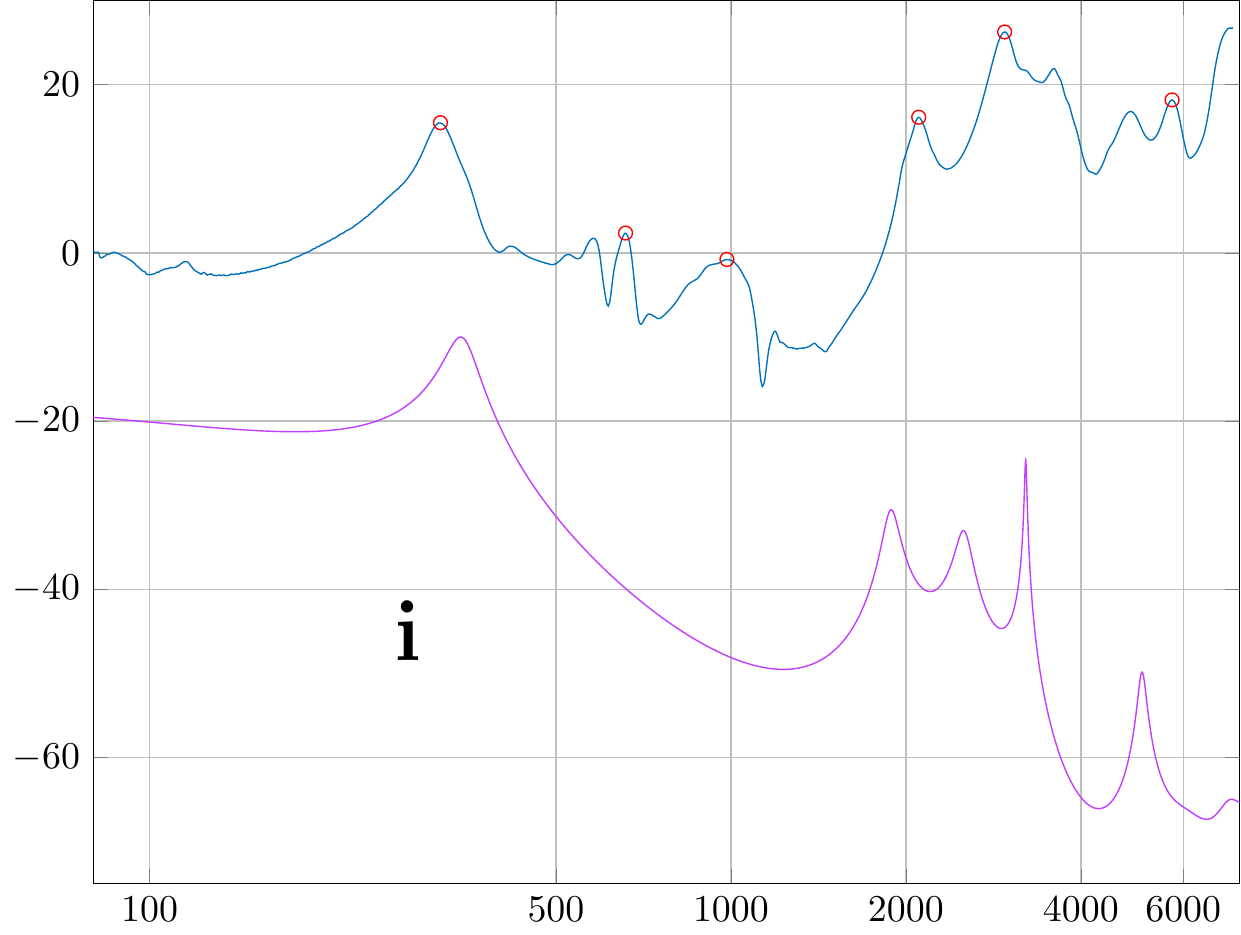}
  \includegraphics[width=0.3\textwidth]{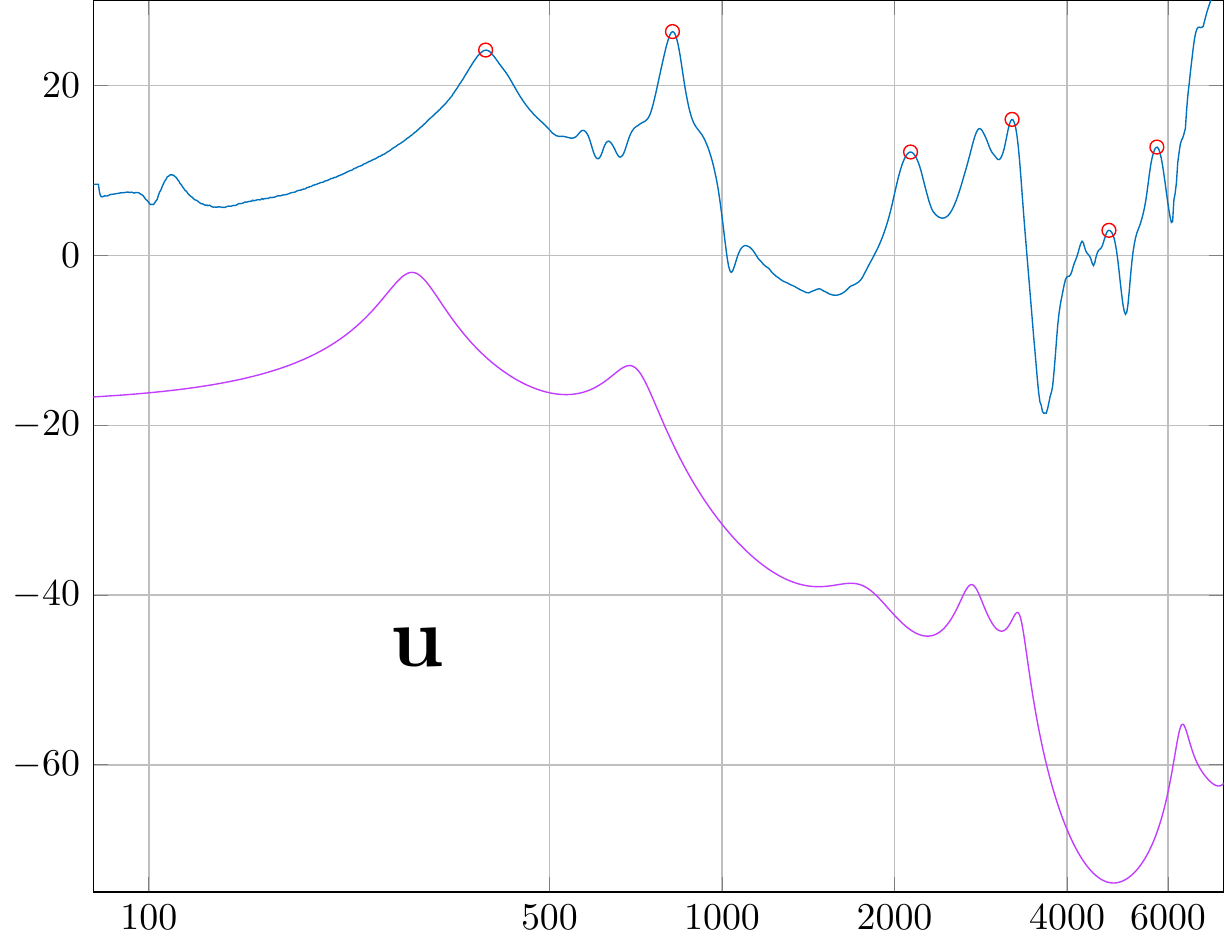}
  \caption{\label{VTResponseFig} The measured frequency amplitude
    response of physical models of VT anatomies corresponding to
    vowels [\textipa{\textscripta, i, u}]. The spectral maxima
    extending to $7350\,\mathrm{Hz}$ were selected so that two peaks
    had to be at least $100\,\mathrm{Hz}$ apart from another with at
    least $4\,\mathrm{dB}$ peak prominence have been marked with
    circles. The lower curves are power spectral envelopes extracted
    from the vowel utterances of the same test subject, recorded in
    the anechoic chamber. }
\end{figure}

The power spectral density is obtained from VT physical models by the
sweep measurements. The sweep is constructed as described in
Section~\ref{CompensationSubSec} to obtain a constant sound pressure
at the output of the source when terminated to the dummy load. The
signal from the measurement microphone at the mouth position of the
physical model is then transformed to an amplitude envelope (similar
approach can be found in~\cite[Fig. 2]{Wolfe:EMS:2016}) by an envelope
detector (i.e., computing a moving average of the nonnegative signal
amplitude). Finally, this output envelope is divided by the similar
envelope from the reference microphone at the source output. The
resulting amplitude envelopes are shown in the top curves of
Fig.~\ref{VTResponseFig}, and the resonance data is given in
Table~\ref{sweepFormantTable}.

\begin{table}[h]
 \centerline{
    \begin{tabular}{|l|c|c|c|c|c|c|}
\hline
 & $P_1$  & $P_2$  & $P_3$ &$P_4$ &$P_5$&$P_6$ \\
\hline
\textbf{[a]} & 635 *  & 1104 *  & 2364 * & 3167 & 4038& X \\
\textbf{[i]} &  316 * & 658 & 984 & 2104 * & 2957 * & 5740 \\
\textbf{[u]}  & 386 *  &819 *  &2132 *  &3206  &4732&5736 \\
\hline
\end{tabular}}
\caption{\label{sweepFormantTable} Peak frequency positions from sweep
  measurements on 3D printed VT physical models. The peaks
  corresponding to the three lowest formants $F_1, F_2$, and $F_3$
  are denoted by an asterisk.}
\end{table}

%% \begin{table}[h] \centerline{ \begin{tabular}{|l|c|c|c|c|c|c|}
%%  \hline & $P_1$ & $P_2$ & $P_3$ &$P_4$ &$P_5$&$P_6$ \\ \hline
%%  \textbf{[a]} & 635&1104&2364&3167 & 4038& X \\ \textbf{[i]} & 316&
%%  658&984&2104 &2957 &5740 \\ \textbf{[u]} &
%%  386&819&2132&3206&4732&5736
%%  \\ \hline \end{tabular}} \caption{\label{sweepFormantTable}
%%  } \end{table}

\subsection{\label{GlottalPulseReconSec} Glottal pulse reconstruction}

\begin{figure}[t]
  \begin{center}
    \begin{subfigure}[b]{0.485\textwidth}
      \centering
      \includegraphics[width=0.49\textwidth]{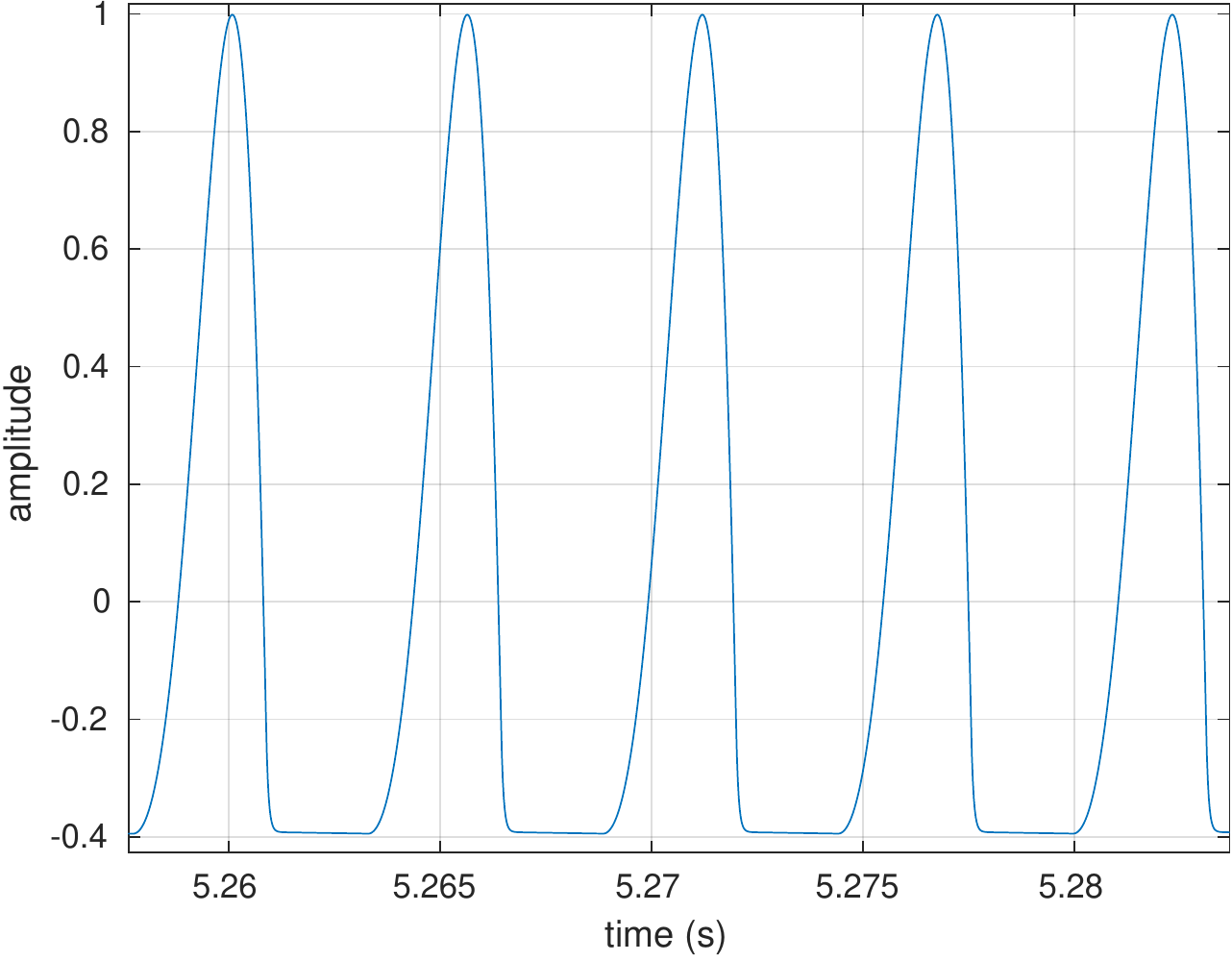}
      \includegraphics[width=0.49\textwidth]{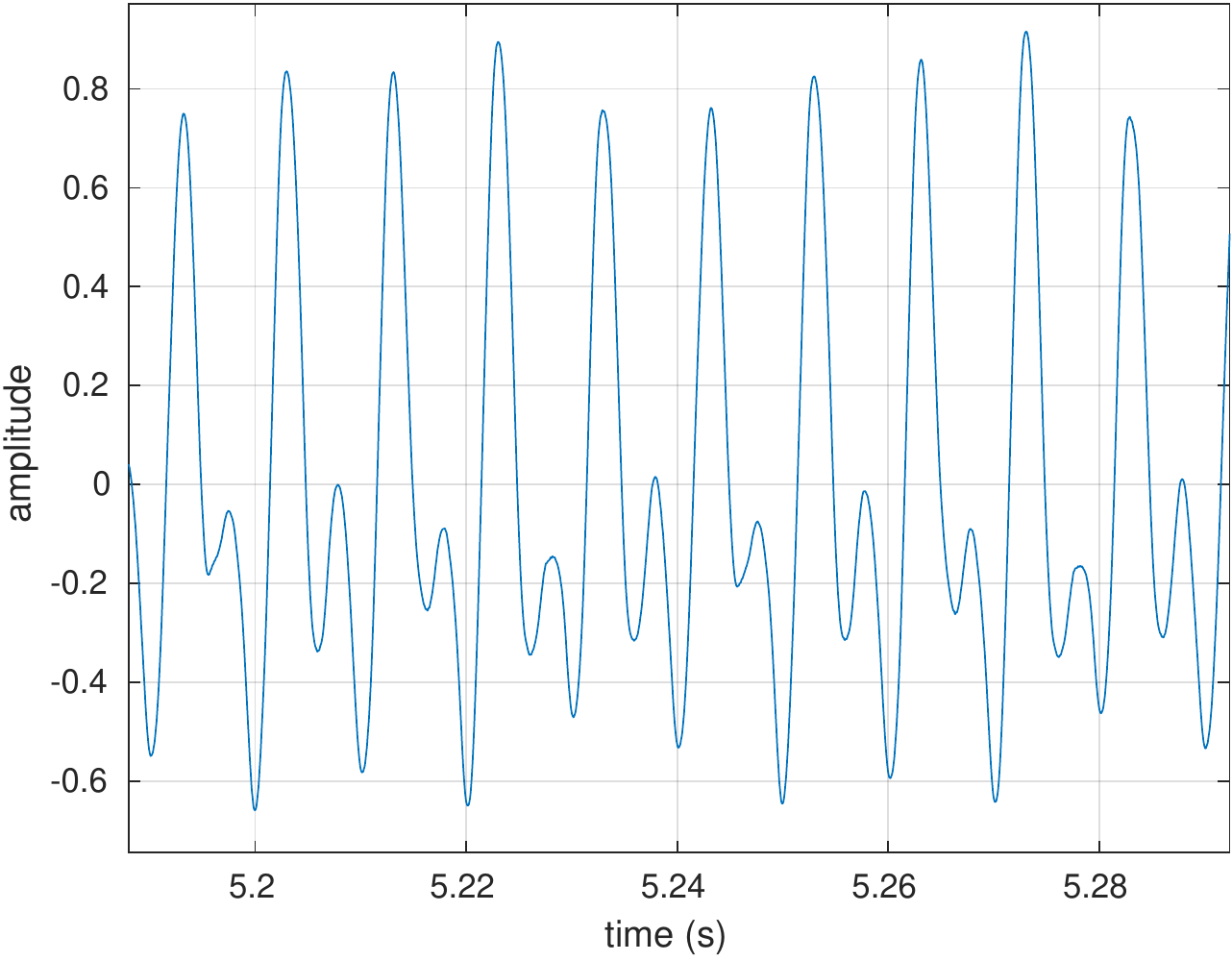}
      \caption{The LF waveform input to the measurement system (left)
        and the corresponding pressure output at the glottal position
        (right).}
        \label{fig:direct}
      \end{subfigure}
      \hspace{0.05cm}
      \begin{subfigure}[b]{0.485\textwidth}
        \centering
        \includegraphics[width=0.49\textwidth]{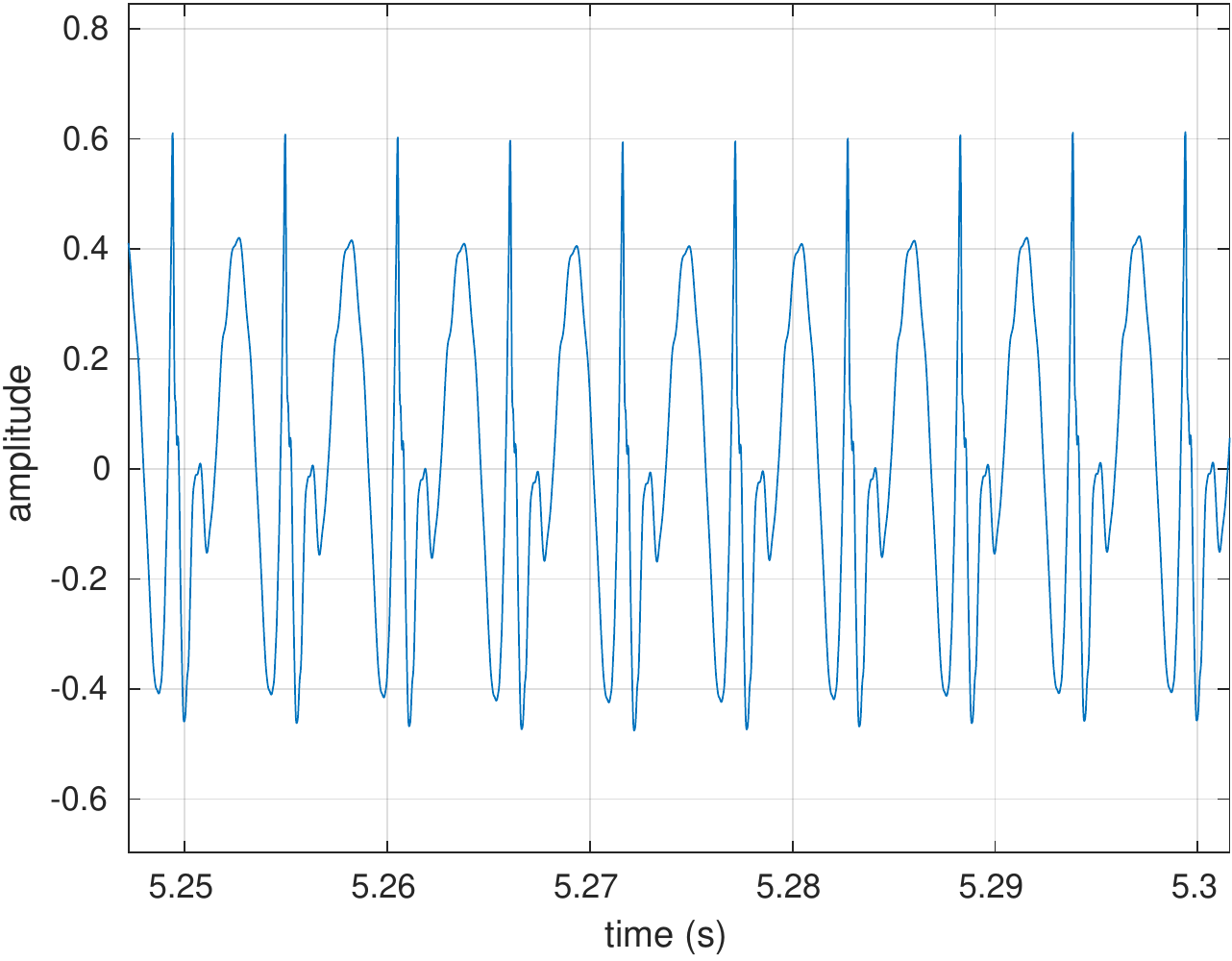}
        \includegraphics[width=0.485\textwidth]{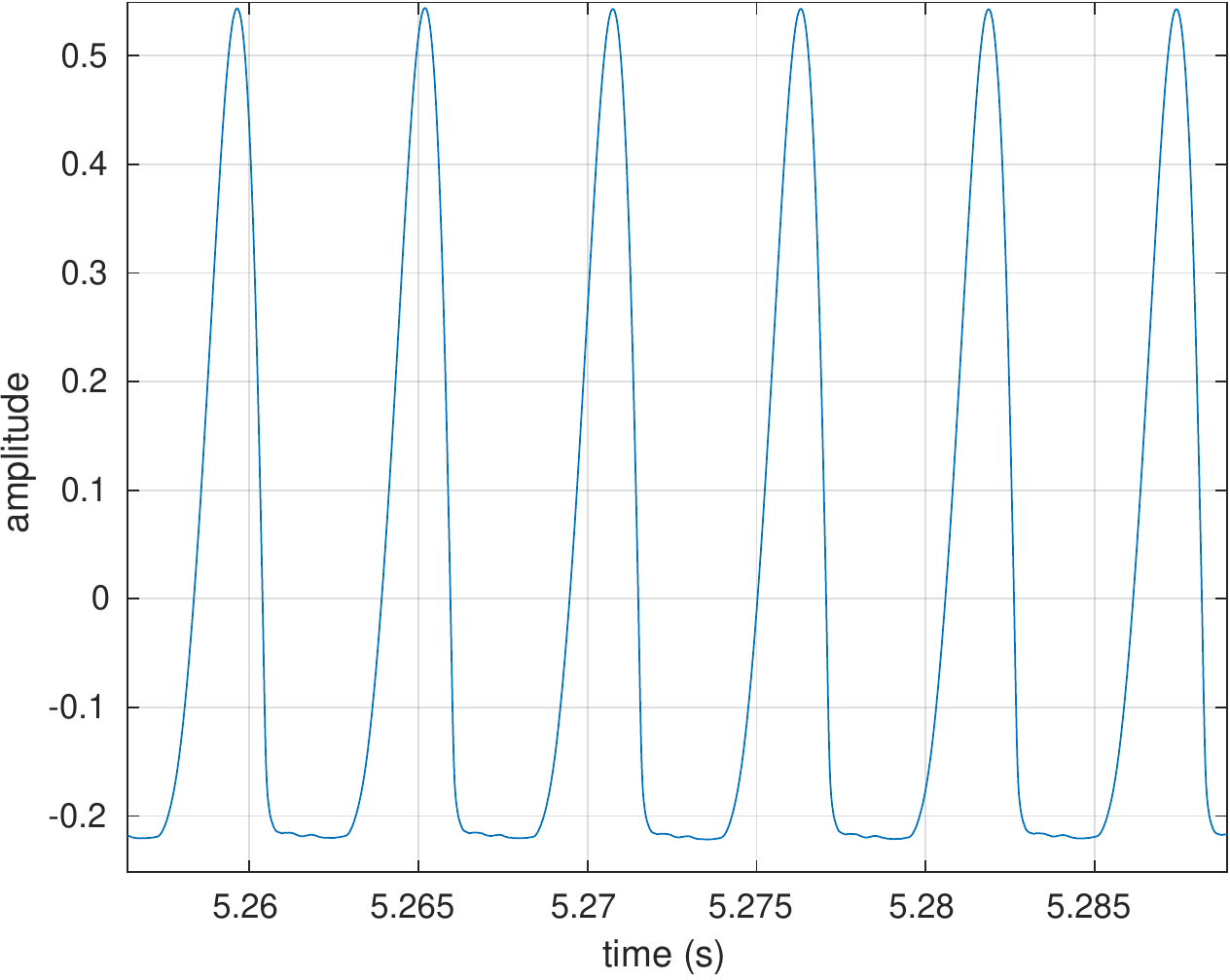}
        \caption{The input waveform produced by regularised
          deconvolution (left) and the corresponding output,
          replicating the LF waveform (right).}
        \label{fig:deconv}
    \end{subfigure}
    \caption{Output waveform reconstruction at $180 \, \mathrm{Hz}$
      using measured impulse response and regularised
      deconvolution.}\label{ReconstructedWaveformFig}
    \end{center}
\end{figure}

The second goal is to reconstruct acoustically reasonable pressure
waveforms at the source output as observed by the reference
microphone.  For reproducing nonsinusoidal target signals, a general
method described in Sections~\ref{DeConvSec}~and~\ref{ImpulseSubSec}
is used to track them. We use the LF waveform shown in
Fig.~\ref{fig:direct} (left panel) as the target signal since it
models the action of the vocal folds during phonation. The regularised
convolution is successful in producing the desired tracking as can be
seen in Fig.~\ref{fig:deconv} (right panel).  For results shown in
Fig.~\ref{ReconstructedWaveformFig}, the impulse response and all
signals have been measured with the source terminated to the vowel
geometry [\textipa{\textscripta}].

\section{Discussion}

After many design cycles for improvements, the proposed acoustic
glottal source appears well suited for its intended use. We now
proceed to discuss remaining shortcomings and possible improvements
for the design and algorithms.

The three most serious shortcomings in the final design are
\textrm{(i)} high TL in the impedance matching cavity due to
attenuation by polyester fibre, \textrm{(ii)} acoustic leakage through
the source chassis, and \textrm{(iii)} the usable low frequency limit
at $\approx 80 \, \mathrm{Hz}$. Since the proposed design is scalable,
the latter two deficiencies are easiest treated by increasing the
physical dimensions, chassis wall thickness, and, hence, the mass of
the source. Using $6''$ or even $8''$ loudspeaker unit with lower bass
resonant frequencies could be considered, equipped with separate
concentric tweeters for producing the higher frequencies. Overly
increasing the size of the source makes it, however, impractical for
demonstration purposes.

Transversal resonances were checked by adding polyester fibre to the
wide parts of the impedance matching cavity which results in a marked
increase in TL of the source. Considering the amplitude response
dynamics of $\approx 35 \, \mathrm{dB}$ of the source shown in
Fig.~\ref{EnvelopeResidualFig}, the output volume remains relatively
low in uniform amplitude sweeps that are produced as explained in
Section~\ref{CompensationSubSec}. Even though the VT physical models
have additional TL of order $20 \ldots 40 \, \mathrm{dB}$ depending on
the vowel and test subject, it is possible to carry out frequency
response of formant position measurements without an anechoic chamber
or a high quality measurement microphone at the mouth position, and
the results are quite satisfactory; see
\cite[Fig.~5]{K-M-O:PPSRDMRI}. To obtain the high quality frequency
response data or carry out waveform reconstructions presented in
Section~\ref{MeasSigSec}, one has to do the utmost to reduce acoustic
leakage, hum, and noise level, including using of the Br\"uel \&
Kj\ae{}ll measurement microphone in the anechoic chamber.  Then
secondary error components emerge as can be observed, e.g., as
roughness between the formant peaks in Fig.~\ref{VTResponseFig}. We
point out that the quality of the microphone used at the mouth opening
does not affect the measured frequencies of the formant
peaks. However, the microphone position or the paraboloid concentrator
shown in \cite[Fig.~4]{K-M-O:PPSRDMRI} does have a small yet
observable effect, in particular, on the lowest resonance frequency of
the physical model.

An attractive way of getting a louder sound source is to use
\emph{Smith slits} \cite{Smith:1953,Dodd:2009} for checking the transversal
resonances within the wide part of the impedance matching cavity.  The
required design work is best carried out using computational design
optimisation methods introduced in
\cite{U-W-B:OVMAH,Y-W-B:LOTSHMIFFDPAH}.

This article does not concern impedance measurements rather than
response between two acoustic pressures. For impedance measurements,
the perturbation velocity should be measured at the output of the
source for which a number of approaches, based on microphones, have
been proposed \cite{Wolfe:AIS:2000, Wolfe:IPM:2006,Wolfe:EEI:2013}. In
the current design, hot wire anemometry at the reference microphone
position would be most suitable; see \cite{Pratt:MAI:1977,
  Kob:MMV:2002}. Even the smallest Microflown unit (see
\cite{Eerden:EWN:1998,Bree:TMN:1996,Bree:TMF:1997}) commercially
available, placed in the middle of the source output channel of
diameter $6\, \mathrm{mm}$, would cause severe back reflections.

We have used two different response compensation techniques in
Section~\ref{CalibrationSec}: weighting for sweeps and regularised
deconvolution for more complicated signals. Using deconvolution for
producing sweeps tens of seconds long is not a practical since the
dimension of Eqs.~\eqref{DiscretisedRegConv} would be too high.  As
opposed to weighted sweeps, regularised deconvolution takes into
account the phase response of the full measurement system.  The
deconvolution is a linear operation whereas the measurement system
shows signs of amplitude nonlinearity in
Fig.~\ref{EnvelopeResidualFig}. This is one of the reasons why
tracking more challenging targets than the LF waveform (e.g., the ramp
signal) will not give as good an outcome. The compensation weight
reconstruction method in Section~\ref{CompensationSubSec} does not
rely on linearity at all, and its performance can be improved by
increasing the sweep length.

One of the challenging secondary objectives is to design dummy loads
of \emph{reasonable physical size} for the source that would present a
constant resistive load over a wide range of frequencies. The dummy
load shown in Fig.~\ref{TractrixHornFig} (right panel) consists of a
tractrix horn tightly filled with polyester fibre, and it has the
property of not being resonant to an observable degree. Two
particularly inspiring examples on the construction of resistive
acoustic loads are given in \cite{Wolfe:AIS:2000} ($42 \,\mathrm{m}$
of insulated steel pipe of inner diam. $7.8 \, \mathrm{mm}$) and
\cite{Wolfe:IPM:2006} ($97 \,\mathrm{m}$ of straight PVC pipe of inner
diam. $15 \, \mathrm{mm}$). The practical challenges in such
approaches are considerable.

We conclude by discussing the numerical efficiency of the discretised
deconvolution proposed in Section~\ref{ImpulseSubSec}. In order to
obtain an $N \log{N}$ algorithm, the $N \times N$ matrices $R$ and $H$
were forced to be circulant. Another way to proceed is allowing $R$ to
be the usual tridiagonal, symmetric, second order difference matrix, 
and $H$ to be the upper triangular matrix obtained from the impulse
response, both noncirculant Toeplitz matrices.  Then the symmetric
matrix $\gamma \left (\kappa I + R^{T}R \right ) + H^{T}H$ in
Eq.~\eqref{DiscretisedRegConv} is a slightly perturbed Toeplitz
matrix, and the required (approximate) solution of the linear system
can be carried out by Toeplitz-preconditioned Conjugate Gradients at
superlinear convergence speed; see, e.g., \cite{JM:PIT}.  Again, an $N
\log{N}$ algorithm is obtained if the matrix/vector products are
implemented by FFT.

\section{Conclusions}

A sound source was proposed for acoustic measurements of vocal tract
physical models, produced by Fast Prototyping methods from Magnetic
Resonance Images. The source design requires only commonly available
components and instruments, and it can be scaled to different
frequency ranges.  Heuristic and numerical methods were used to
understand and to optimise the source design and performance. Two
kinds of algorithms were proposed for compensating the source
nonoptimality: (\textrm{i}) an iterative process for producing uniform
amplitude sound pressure sweeps, and (\textrm{ii}) a method based on
regularised deconvolution for replicating target sound pressure
waveforms at the source output.  The sound source together with the
two compensation algorithms, written in MATLAB code, was deemed
successful based on measurements on the vocal tract geometry
corresponding to vowel [\textipa{\textscripta}] of a male speaker.

\section*{Acknowledgments}

 The authors wish to thank for consultation and facilities
 Dept. Signal Processing and Acoustics, Aalto University
 (Prof.~P.~Alku, Lab.~Eng.~I.~Huhtakallio, M.~Sc.~M.~Airaksinen) and
 Digital Design Laboratory, Aalto University (M.~Arch.~A.~Mohite).

 The authors have received financial support from Instrumentarium
 Science Foundation, Magnus Ehrnrooth Foundation, Niilo Helander
 Foundation, and Vilho, Yrj\"o and Kalle V\"ais\"al\"a Foundation.

\section*{References}
\bibliography{GsourceLocal}
%\bibliography{../../bibtex/ac-flow,../../bibtex/master_new,../../bibtex/jmalinen}
%\bibliography{../../bibtex/master_new,../../bibtex/jmalinen}

\end{document}